\input harvmac
\noblackbox
\newcount\fignoa
\figno=0
\def\fig#1#2#3{
\par\begingroup\parindent=0pt\leftskip=1cm\rightskip=1cm\parindent=0pt
\baselineskip=11pt \global\advance\figno by 1 \midinsert
\epsfxsize=#3 \centerline{\epsfbox{#2}} \vskip 12pt {\bf Fig.
\the\figno:} #1\par
\endinsert\endgroup\par
}
\def\figlabel#1{\xdef#1{\the\figno}}
\def\encadremath#1{\vbox{\hrule\hbox{\vrule\kern8pt\vbox{\kern8pt
\hbox{$\displaystyle #1$}\kern8pt} \kern8pt\vrule}\hrule}}

\def\apm{{\alpha^\prime}}
\def\bba{{\bf{A}}}

\def\bbm{{\bf{M}}}

\def\bbs{{\bf{S}}}



\def\unlockat{\catcode`\@=11}
\def\lockat{\catcode`\@=12}

\unlockat

\def\newsec#1{\global\advance\secno by1\message{(\the\secno. #1)}
\global\subsecno=0\global\subsubsecno=0\eqnres@t\noindent
{\bf\the\secno. #1} \writetoca{{\secsym}
{#1}}\par\nobreak\medskip\nobreak}
\global\newcount\subsecno \global\subsecno=0
\def\subsec#1{\global\advance\subsecno
by1\message{(\secsym\the\subsecno. #1)}
\ifnum\lastpenalty>9000\else\bigbreak\fi\global\subsubsecno=0
\noindent{\it\secsym\the\subsecno. #1} \writetoca{\string\quad
{\secsym\the\subsecno.} {#1}}
\par\nobreak\medskip\nobreak}
\global\newcount\subsubsecno \global\subsubsecno=0
\def\subsubsec#1{\global\advance\subsubsecno by1
\message{(\secsym\the\subsecno.\the\subsubsecno. #1)}
\ifnum\lastpenalty>9000\else\bigbreak\fi
\noindent\quad{\secsym\the\subsecno.\the\subsubsecno.}{#1}
\writetoca{\string\qquad{\secsym\the\subsecno.\the\subsubsecno.}{#1}}
\par\nobreak\medskip\nobreak}

\def\subsubseclab#1{\DefWarn#1\xdef
#1{\noexpand\hyperref{}{subsubsection}%
{\secsym\the\subsecno.\the\subsubsecno}%
{\secsym\the\subsecno.\the\subsubsecno}}%
\writedef{#1\leftbracket#1}\wrlabeL{#1=#1}}
\lockat

\overfullrule=0pt


%

\def\tilde{\widetilde}
\def\bar{\overline}

%
\def\inbar{\,\vrule height1.5ex width.4pt depth0pt}
\def\IB{\relax{\rm I\kern-.18em B}}
\def\IC{\relax\hbox{$\inbar\kern-.3em{\rm C}$}}
\def\ID{\relax{\rm I\kern-.18em D}}
\def\IE{\relax{\rm I\kern-.18em E}}
\def\IF{\relax{\rm I\kern-.18em F}}
\def\IG{\relax\hbox{$\inbar\kern-.3em{\rm G}$}}
\def\IH{\relax{\rm I\kern-.18em H}}
\def\II{\relax{\rm I\kern-.18em I}}
\def\IK{\relax{\rm I\kern-.18em K}}
\def\IL{\relax{\rm I\kern-.18em L}}
\def\IM{\relax{\rm I\kern-.18em M}}
\def\IN{\relax{\rm I\kern-.18em N}}
\def\IO{\relax\hbox{$\inbar\kern-.3em{\rm O}$}}
\def\IP{\relax{\rm I\kern-.18em P}}
\def\IQ{\relax\hbox{$\inbar\kern-.3em{\rm Q}$}}
\def\IR{\relax{\rm I\kern-.18em R}}
\font\cmss=cmss10 \font\cmsss=cmss10 at 7pt
\def\IZ{\relax\ifmmode\mathchoice
{\hbox{\cmss Z\kern-.4em Z}}{\hbox{\cmss Z\kern-.4em Z}}
{\lower.9pt\hbox{\cmsss Z\kern-.4em Z}} {\lower1.2pt\hbox{\cmsss
Z\kern-.4em Z}}\else{\cmss Z\kern-.4em Z}\fi}
\def\IGa{\relax\hbox{${\rm I}\kern-.18em\Gamma$}}
\def\IPi{\relax\hbox{${\rm I}\kern-.18em\Pi$}}
\def\ITh{\relax\hbox{$\inbar\kern-.3em\Theta$}}
\def\IOm{\relax\hbox{$\inbar\kern-3.00pt\Omega$}}

\def\pf{{\rm Pf ~}}
\font\zfont = cmss10 
 
\def\bigone{\hbox{1\kern -.23em {\rm l}}}
\def\ZZ{\hbox{\zfont Z\kern-.4emZ}}

\def\half{{\litfont {1 \over 2}}}

\def\CM{{\cal M}}

\def\G{\Gamma}
\def\a{\alpha}

\def\e{\epsilon}

\def\th{\theta}

\def\i{\iota}
\def\k{\kappa}

\def\k{\kappa}

\def\IR{\relax{\rm I\kern-.18em R}}
\def\I1{\relax{\rm I\kern-.6em 1}}
\def\Dsl{\,\raise.15ex\hbox{/}\mkern-13.5mu D}
\def\Gsl{\,\raise.15ex\hbox{/}\mkern-13.5mu G}
\def\Csl{\,\raise.15ex\hbox{/}\mkern-13.5mu C}
\font\cmss=cmss10 \font\cmsss=cmss10 at 7pt

\def\pf{{\rm Pf ~}}
\font\zfont = cmss10 
 
\def\bigone{\hbox{1\kern -.23em {\rm l}}}
\def\ZZ{\hbox{\zfont Z\kern-.4emZ}}
\def\half{{\litfont {1 \over 2}}}

\def\CM{{\cal M}}

\def\IL{\relax{\rm I\kern-.18em L}}
\def\IH{\relax{\rm I\kern-.18em H}}
\def\IR{\relax{\rm I\kern-.18em R}}
\def\IC{\relax\hbox{$\inbar\kern-.3em{\rm C}$}}
\def\IZ{\relax\ifmmode\mathchoice
{\hbox{\cmss Z\kern-.4em Z}}{\hbox{\cmss Z\kern-.4em Z}}
{\lower.9pt\hbox{\cmsss Z\kern-.4em Z}} {\lower1.2pt\hbox{\cmsss
Z\kern-.4em Z}}\else{\cmss Z\kern-.4em Z}\fi}
\def\CM {{\cal M}}

\def\CF {{\cal F}}

\def\CL {{\cal L}}

\def\CM {{\cal M}}

\def\kb{{\bar k}}

\font\manual=manfnt \def\dbend{\lower3.5pt\hbox{\manual\char127}}

\def\IZ{\relax\ifmmode\mathchoice
{\hbox{\cmss Z\kern-.4em Z}}{\hbox{\cmss Z\kern-.4em Z}}
{\lower.9pt\hbox{\cmsss Z\kern-.4em Z}} {\lower1.2pt\hbox{\cmsss
Z\kern-.4em Z}}\else{\cmss Z\kern-.4em Z}\fi}
\def\half {{1\over 2}}

\def\p{\partial}

\def\CM {{\cal M}}


\def\IZ{\relax\ifmmode\mathchoice
{\hbox{\cmss Z\kern-.4em Z}}{\hbox{\cmss Z\kern-.4em Z}}
{\lower.9pt\hbox{\cmsss Z\kern-.4em Z}} {\lower1.2pt\hbox{\cmsss
Z\kern-.4em Z}}\else{\cmss Z\kern-.4em Z}\fi}
\def\IB{\relax{\rm I\kern-.18em B}}
\def\IC{{\relax\hbox{$\inbar\kern-.3em{\rm C}$}}}
\def\ID{\relax{\rm I\kern-.18em D}}
\def\IE{\relax{\rm I\kern-.18em E}}
\def\IF{\relax{\rm I\kern-.18em F}}
\def\IG{\relax\hbox{$\inbar\kern-.3em{\rm G}$}}
\def\IGa{\relax\hbox{${\rm I}\kern-.18em\Gamma$}}
\def\IH{\relax{\rm I\kern-.18em H}}
\def\II{\relax{\rm I\kern-.18em I}}
\def\IK{\relax{\rm I\kern-.18em K}}
\def\IP{\relax{\rm I\kern-.18em P}}

\def\IQ{\relax\hbox{$\inbar\kern-.3em{\rm Q}$}}
\def\IP{\relax{\rm I\kern-.18em P}}
\def\I1{\relax{\rm 1\kern-.18em I}}

\def\inbar{\,\vrule height1.5ex width.4pt depth0pt}

\def\p{\partial}

\font\cmss=cmss10 \font\cmsss=cmss10 at 7pt
\def\IR{\relax{\rm I\kern-.18em R}}

\def\Tr{{{\rm Tr}}}
\def\vol{{\rm vol}}


\def\boxit#1{\vbox{\hrule\hbox{\vrule\kern8pt
\vbox{\hbox{\kern8pt}\hbox{\vbox{#1}}\hbox{\kern8pt}}
\kern8pt\vrule}\hrule}}
\def\mathboxit#1{\vbox{\hrule\hbox{\vrule\kern8pt\vbox{\kern8pt
\hbox{$\displaystyle #1$}\kern8pt}\kern8pt\vrule}\hrule}}


\def\inbar{\,\vrule height1.5ex width.4pt depth0pt}

\def\p{\partial}

\font\cmss=cmss10 \font\cmsss=cmss10 at 7pt
\def\IR{\relax{\rm I\kern-.18em R}}

\def\vol{{\rm vol}}

\def\tanh{{\rm tanh}}
\def\mb{{\bar{m}}}
\def\nb{{\bar{n}}}

\lref\bandequiv{I. Bandos, K. Lechner, A. Nurmagambetov, P. Pasti,
D. Sorokin, and M. Tonin, ``On the equivalence of different
formulations of the M-Theory Five-Brane,'' hep-th/9703127}

\lref\cw{M.~Cederwall, A.~von Gussich, B.E.~Nilsson, P.~Sundell
and A.~Westerberg, ``The Dirichlet super-p-branes in
ten-dimensional type IIA and IIB supergravity,'' Nucl.\ Phys.\
{\bf B490}, 179 (1997) hep-th/9611159.} \lref\bt{E.~Bergshoeff and
P.K.~Townsend, ``Super D-branes,'' Nucl.\ Phys.\ {\bf B490}, 145
(1997) hep-th/9611173.} \lref\bkop{E.~Bergshoeff, R.~Kallosh,
T.~Ortin and G.~Papadopoulos, ``Kappa-symmetry, supersymmetry and
intersecting branes,'' Nucl.\ Phys.\ {\bf B502}, 149 (1997)
hep-th/9705040.} \lref\bbs{K.~Becker, M.~Becker and A.~Strominger,
``Five-branes, membranes and nonperturbative string theory,''
Nucl.\ Phys.\ {\bf B456}, 130 (1995) hep-th/9507158.}

\lref\cds{A. Connes, M. Douglas, and A. Schwarz, ``Noncommutative
geometry and matrix theory: compactification on tori,"
hep-th/9711162; JHEP {\bf 9802} (1998) 003.}

\lref\cornalba{ L. Cornalba, ``D-brane Physics and Noncommutative
Yang-Mills Theory,'' hep-th/9909081}

\lref\hsw{P.S. Howe and E. Sezgin, ``D=11, p=5,'' Phys. Lett.
{\bf B 394} (1997) 62, hep-th/9611008;
P.S. Howe, E. Sezgin, and P.C. West, ``Covariant Field Equations
of the M Theory Five-Brane,''
Phys. Lett. {\bf B 399} (1997) 49, hep-th/9702008.}

 \lref\sw{N.~Seiberg and E.~Witten, ``String theory and
noncommutative geometry,'' JHEP {\bf 9909} (1999) 032;
hep-th/9908142.}

\lref\fks{S. Ferrara, R. Kallosh and A. Strominger, ``N=2 extremal
black holes," Phys. Rev. {\bf D52} (1995) 5412; hep-th/9508072.}

\lref\as{A. Strominger, ``Macroscopic entropy of N=2 extremal
black holes," Phys. Lett. {\bf B383} (1996) 39; hep-th/9602111.}

\lref\clash{A.~Losev, G.~Moore, N.~Nekrasov and S.~Shatashvili,
``Chiral Lagrangians, anomalies, supersymmetry, and holomorphy,''
Nucl.\ Phys.\ {\bf B484} (1997) 196;  hep-th/9606082.}

\lref\leung{N.C. Leung, ``Einstein type metrics and stability on
vector bundles," J. Differential Geom. {\bf 45} (1997) 514;
``Symplectic structures in gauge theory," Commun. Math. Phys. {\bf
193} (1998) 47.}

\lref\hl{R. Harvey and H.B. Lawson, Jr., ``Calibrated geometries,"
Acta Math. {\bf 148} (1982) 47.}

 \lref\hm{J. Harvey and G. Moore,
``Superpotentials and membrane instantons," hep-th/9907026.}

\lref\gkv{S. Gukov,
``Solitons, superpotentials and callibrations," hep-th/9911011.}

\lref\figue{B.S.~Acharya, J.M.~Figueroa-O'Farrill, B.~Spence and
M.~O'Loughlin, ``Euclidean D-branes and higher-dimensional gauge
theory,'' Nucl.\ Phys.\ {\bf B514}(1998) 583; hep-th/9707118.}

\lref\bbcia{K.~Becker, M.~Becker, D.R.~Morrison, H.~Ooguri, Y.~Oz
and Z.~Yin, ``Supersymmetric cycles in exceptional holonomy
manifolds and Calabi-Yau 4-folds,'' Nucl.\ Phys.\ {\bf B480}
(1996) 225; hep-th/9608116.}

\lref\gp{G.W. Gibbons and G. Papadopoulos, ``Calibrations and
intersecting branes," Commun. Math. Phys. {\bf 202} (1999) 593;
hep-th/9803163.}

\lref\nekschw{N.~Nekrasov and A.~Schwarz, ``Instantons on
noncommutative $\IR^4$ and (2,0) superconformal six dimensional
theory,'' Commun.\ Math.\ Phys.\ {\bf 198} (1998) 689;
hep-th/9802068.}

\lref\ooy{H.~Ooguri, Y.~Oz and Z.~Yin, ``D-branes on Calabi-Yau
spaces and their mirrors,'' Nucl.\ Phys.\ {\bf B477} (1996) 407;
hep-th/9606112.}

\lref\tseytlinrev{A. Tseytlin, ``Born-Infeld action, supersymmetry
and string theory," hep-th/9908105.}

\lref\kontsevich{M. Kontsevich, ``Deformation quantization of
Poisson manifolds," q-alg/9709040.}

\lref\pst{I.~Bandos, K.~Lechner, A.~Nurmagambetov, P.~Pasti,
D.~Sorokin and M.~Tonin, ``Covariant action for the
super-five-brane of M-theory,'' Phys.\ Rev.\ Lett.\ {\bf 78}, 4332
(1997) hep-th/9701149.}

\lref\wittendinst{E. Witten, ``World-sheet corrections via
D-instantons,'' hep-th/9907041.}

\lref\wittenfiveact{E. Witten, ``Five-brane effective action in
$M$-theory,'' J. Geom. Phys. {\bf 22} (1997) 103; hep-th/9610234.}

\lref\wittenbcov{E. Witten, ``Quantum background independence in
string theory,'' hep-th/9306122  }

\lref\todorov{A. Todorov, ``Witten's geometric quantization of the
moduli of CY threefolds,'' preprint}

\lref\dvvunpub{R. Dijkgraaf, E. Verlinde, and M. Vonk, ``On the 
partition sum of the NS five-brane,'' hep-th/0205281. S.
Shatashvili, unpublished.}

\lref\hswi{P.S. Howe, E. Sezgin, and P.C. West, ``The
six-dimensional self-dual tensor,'' Phys. Lett. {\bf B 400} (1997) 255,
hep-th/9702111}

\lref\bcov{M. Bershadsky, S. Cecotti, H. Ooguri, and C. Vafa,
``Kodaira-Spencer theory of gravity and exact results for quantum
string amplitudes,'' Comm. Math. Phys. {\bf 165} (1994) 311.}

\lref\tian{G. Tian, ``Smoothness of the universal deformation
space of compact Calabi-Yau manifolds and its Petersson-Weil
metric," in {\it Mathematical aspects of string theory}, S.T. Yau,
ed., World Scientific, 1987.}

\lref\todorov{A. Todorov, ``The Weil-Petersson geometry of the
moduli space of ${\rm SU}(n\geq 3)$ (Calabi-Yau) manifolds. I,"
Commun. Math. Phys. {\bf 126} (1989) 325.}

\lref\single{E.~Bergshoeff, M.J.~Duff, C.N.~Pope and E.~Sezgin,
``Supersymmetric Supermembrane Vacua And Singletons,'' Phys.\
Lett.\ {\bf B199}, 69 (1987).}

\lref\yauremark{S.-T. Yau, private communication.}

\lref\hlw{P.S. Howe, N.D. Lambert, P.C. West,
``The Self-Dual String Soliton,''  hep-th/9709014; Nucl. Phys.
{\bf B 515} (1998) 203.}

\lref\glw{J.P. Gauntlett, N.D. Lambert, and P.C. West,
``Supersymmetric Fivebrane Solitons,'' hep-th/9811024;
Adv. Theor. Math. Phys. {\bf 3} (1999) 91.}

\lref\gauntlett{J.P. Gauntlett, ``Membranes on 5branes,''
hep-th/9906162.}

\lref\kmt{
P.~Kaste, R.~Minasian and A.~Tomasiello,
``Supersymmetric M-theory compactifications with fluxes on  seven-manifolds and G-structures,''
JHEP {\bf 0307}, 004 (2003) hep-th/0303127.
}

\lref\kb{
K.~Becker,
``A note on compactifications on Spin(7)-holonomy manifolds,''
JHEP {\bf 0105}, 003 (2001) hep-th/0011114.
}

%

%

\Title{\vbox{\baselineskip12pt \hbox{YCTP-P28-99}
\hbox{hep-th/9911206} }} {\vbox{ \centerline{Nonlinear Instantons}
\centerline{from} \centerline{Supersymmetric $p$-Branes}  }}

\bigskip
\centerline{ Marcos Mari\~no,$^{\spadesuit}$\foot{Address after
Jan. 1, 2000: Dept. of Physics and Astronomy, Rutgers University,
Piscataway, NJ 08855-0849, USA}Ruben
Minasian,$^{\spadesuit}$\foot{Centre de Physique Th{\'e}orique,
Ecole Polytechnique,  F-91128 Palaiseau, France}
 Gregory Moore,$^{\spadesuit} {}^1$ and Andrew Strominger$^{\clubsuit}$}

\bigskip
\centerline{$^\spadesuit$ Department of Physics, Yale University}
\centerline{New Haven, CT 06520, USA}

\centerline{$^\clubsuit$ Jefferson Physical Laboratories, Harvard
University} \centerline{Cambridge, MA 02138, USA}

\bigskip
\centerline{\bf Abstract}
\medskip
\noindent Supersymmetric configurations of type II $D$-branes with
nonzero gauge field strengths in general supersymmetric
backgrounds with nonzero $B$ fields are analyzed using the
$\kappa$-symmetric worldvolume action.  It is found in dimension
four or greater that the usual instanton equation for the gauge
field obtains a nonlinear deformation. The deformation is
parameterized by the topological data of the $B$-field, the
background geometry and the cycle wrapped by the brane. In the
appropriate dimensions, limits and settings these equations reduce
to deformed instanton equations recently found in the context of
noncommutative geometry as well as those following from
Lagrangians based on Bott-Chern forms. We further consider
instantons comprised of M5-branes wrapping a Calabi-Yau space with
non-vanishing three-form field strengths. It is shown that the
instanton equations for the three-form are related to the
Kodaira-Spencer equations.

\Date{November 25, 1999}

%
%
\listtoc \writetoc

\vfill\eject

\newsec{Introduction and Conclusion}

Supersymmetric cycle conditions are of importance in string theory
because these are the equations governing the existence of BPS
states and supersymmetric  instantons associated with wrapped
$p$-branes. These supersymmetric cycle conditions  are equations
for a {\it pair} consisting of an embedded manifold in spacetime,
together with a gauge connection on the embedded manifold. There
is by now a rich literature for the special case in which the
gauge field strength $F$ on the brane as well as the background
supergravity potentials $B$ vanish. The general case of nonzero
$F$ or $B$ in contrast has received less attention.

In this paper the more general case is addressed. $D$-branes with
nonzero $U(1)$ gauge fields $F$ in supersymmetric $IIA$ and $IIB$
supergravity backgrounds with nonzero $B$ fields are analyzed
using the recently discovered $\kappa$-symmetric worldvolume
actions \refs{\cw,\bt,\bkop}. In all cases considered, we find
that the conditions governing supersymmetrically embedded cycles
are unchanged by $B\not=0$.\foot{The cases we consider do not
include certain interesting singularities, such as occur in
BIons.} In contrast, the equation for the gauge field on the cycle
is, in dimension four or greater, an intriguing nonlinear
deformation of the usual instanton equation. The deformed equation
depends on the topological data of both the cycle and the $B$
field.

An important motivation for this work is its application to the
study of Yang-Mills theory in a non-commutative geometry. The
connection between noncommutative geometry and string theory was
first noted by Connes, Douglas, and Schwarz \cds. Recently,
Seiberg and Witten \sw\ have related the study of BPS
configurations of D-branes to various aspects of noncommutative
geometry. In particular it is shown how the presence of a nonzero
$B$-field deforms the instanton equations, in a way related to the
noncommutative instanton equations of Nekrasov and Schwarz
\nekschw. We will find that our deformed equations reduce to those
of \sw\ in the appropriate limit and setting. It further
generalizes those equations to curved backgrounds and to higher
dimensions. In $\IR^4$ we find an additional deformation parameter
away from the limit considered in \sw.\foot{This additional
parameter was known to the authors of \sw. } In particular we find
that the Hermitian Yang-Mills equations of six dimensions are also
modified, and that the discussion of \sw\ relating BPS conditions
to the noncommutative instanton equations of Nekrasov-Schwarz
\nekschw\ generalizes nicely  to this case. We also find that the
$G_2$ and ${\rm Spin}(7)$ instanton equations are not deformed,
and the instanton equations on calibrated 4-cycles in such
manifolds are likewise not deformed.

This paper also has two unexpected results.  The first novelty is
that the nonlinear deformations of the instanton equations arising
from BPS conditions are closely related to the equations for
hermitian metrics on holomorphic vector bundles discussed by Leung
\leung\ and by Losev et. al. \clash. The second novelty  appears
when we further extend our analysis to the case of an M5-brane
wrapping a Calabi-Yau threefold.  The M5-brane worldvolume
contains a closed rank three antisymmetric tensor field strength
$H$ which obeys a nonlinear ``self-duality constraint'' \hsw\pst.
In \hswi\ a nonlinear change of variables to a 3-form $h$ was
found such that $h$ is self-dual, but satisfies a nonlinear
equation of motion. We show that, when combined with the condition
of preserving a supersymmetry,  this nonlinear equation of motion
is just the Kodaira-Spencer equation together with a gauge
condition. The gauge condition is a deformation of the standard
one, and it shown that a solution exists through third order in
perturbation theory.

This paper raises a number of interesting open problems. In
principle non-abelian versions of our deformed equations could be
derived via string theory by considering multiply-wrapped
$D$-branes. In practice this would be difficult. However in most
cases the equations we write have obvious non-abelian
generalizations which we expect to apply to this case. Moreover,
the  relation to \clash\ opens up a host of new issues and
questions related to the possible use of higher dimensional ``$bc$
systems.''

A brief outline of the paper is the following.
 Section 2 is a
lightning review of the relevant $\kappa$-symmetric $D$-branes
\refs{\cw,\bt}. In section 3 we show that the cycle embeddings are
undeformed and derive the deformed instanton equations for a
variety of branes embedded in manifolds of $SU(2)$, $SU(3)$,
$G_2$, $SU(4)$ and ${\rm Spin(7)}$ holonomy. A group theoretic
analysis of the equations and their solutions for $\IR^4$, $\IR^6$
and $\IR^8$ is given in section 4. In section 5 we relate a limit
of the $\IR^4$ equations to those derived by Seiberg and Witten in
the context of noncommutative geometry, and then generalize the
relation to $\IR^6$. In section 7, we briefly discuss the relation
of the present results to those of Leung \leung\ and Losev et. al.
\clash. Finally, in section 8, we analyze the supersymmetry
conditions for an M5 brane wrapping a Calabi-Yau threefold.

\newsec{Review of $\k$-symmetry and the $\G$-operator}

We start with a very brief review of D-brane actions,
$\k$-symmetry and the properties of the $\G$-operator that are
crucial for our discussion. The action was constructed in
\refs{\cw, \bt} and is of the form (for constant dilaton)
\eqn\action{ I_p = I_{\rm DBI} + I_{\rm WZ} = -T_p \int_W
d^{p+1}\sigma \, \sqrt {{\rm det} (g_{\mu\nu}+{M}_{\mu\nu})} + \mu
\int_W C \wedge e^{{M}} . }
Here $W$ is the brane worldvolume. $T_p$ and $\mu$ are the brane
tensions and charges, respectively. The metric on $W$
\eqn\metric{ g_{\mu \nu} = E_\mu{}^aE_\nu{}^b\eta_{ab}\, , }
is induced by supermaps $Z=(X, \th)$ and super-vielbein
\eqn\sv{ E_\mu{}^{\bba} = \partial_\mu Z^{\bbm}
E_{\bbm}{}^{\bba}\, .}
The frame index $\bba$ decomposes into 10 vector ($a$) and 32
spinor ($\a$) indices (and an $SL(2, \IR)$ index ($A$) in type IIb
case). $M_{\mu \nu}\ (\mu=1,\cdots (p+1))$ is the modified 2-form
field strength $M= 2\pi \alpha'(F+B)$ with $B$ being the pull-back
of the NS two-form field to the worldvolume. The couplings to the
background RR fields are given by the second term in \action,
where $$ C = \sum_{r=0}^{10} C^{(r)} $$ is a formal sum of the RR
fields
 $C^{(r)}$ (we are ignoring the gravitational
couplings here).

Since in addition to $\k$-symmetry, the classical D-brane actions
have spacetime supersymmetry, we can combine both, and in
particular determine the fraction of unbroken supersymmetry by the
dimension of the solution space  of the equation
\refs{\bbs,\single}:
\eqn\proj{(1-\G)\eta =0,}
where $\eta$ is the spacetime supersymmetry parameter, and $\G$ is
an Hermitian traceless matrix: \eqn\gam{ {\rm tr}\ \Gamma=0,\,
\,\,\,\,\,\,\,\, \Gamma^2=1\, .}

The explicit form  for $\G$ will be important for our analysis
(note that we are working with Euclidean branes):
\eqn\gamm{ \Gamma={{\sqrt {|g|} \over \sqrt{|g + {M}|}}}
\sum^{\infty}_{n=0} {1\over 2^n n!} \gamma^{\mu_1\nu_1\dots
\mu_n\nu_n} {M}_{\mu_1\nu_1}\dots {M}_{\mu_n\nu_n} J^{(n)}_{(p)}\,
,}
where $\vert g \vert := {\rm det}( g_{\mu \nu})$, $\vert
g+{M}\vert :={\rm det} (g_{\mu\nu }+ {M}_{\mu\nu})$, and
\eqn\defs{ J^{(n)}_{(p)}=\cases{i(\Gamma_{11}\big)^{n+{p-2\over
2}} \Gamma_{(0)}
\,\,\,\,\,\,\,\,\,\,\,\,\,\,\,\,\,\,\,\,\,\,\,\,\,\,\, ({\rm for
\,\,\,\, IIA}) \cr (-1)^n(\sigma_3)^{n+{p-3 \over 2}}
\sigma_2\otimes \Gamma_{(0)}\,\,\,\,\,\,\, ({\rm for \,\,\,\,
IIB}). }}
Finally, $\Gamma_{(0)}$ is defined as
\eqn\gnot{ \Gamma_{(0)}={{1 \over (p+1)! \sqrt{|g|}}}
\epsilon^{\mu_1\dots \mu_{(p+1)}} \gamma_{\mu_1\dots \mu_{(p+1)}},
}
with $$ \gamma_\mu = E_\mu{}^a \Gamma_a\, , $$ where $\{\Gamma_a;
a=0,\dots 9\}$ are the spacetime gamma-matrices.

A very important feature is that the non-linear dependence on $M$
can be expressed in the form \bkop:
\eqn\gammaaa{ \Gamma= e^{-a/2}\ \Gamma_{(0)}^{\prime}\ e^{a/2}\, ,
}
where $a = a({M})$ contains all the dependence on $M$ and
$\Gamma^{\prime}_{(0)}$ (which depends only on $X$) is also an
Hermitian traceless matrix (i.e.~tr $\Gamma_{(0)}^{\prime} = 0$
and $\left(\Gamma_{(0)}^{\prime}\right)^2 = 1$) given by:
\eqn\gpnot{\Gamma^{\prime}_{(0)}=\cases{(\Gamma_{11})^{p-2\over 2}
\Gamma_{(0)} \,\,\,\,\,\,\,\,\,\,\,\,\,\,\,\,\, ({\rm for \,\,\,\,
IIA}) \cr (\sigma_3)^{p-3\over2} \sigma_2\otimes
\Gamma_{(0)}\,\,\,\,\,\,\, ({\rm for \,\,\,\, IIB}). }}
An explicit expression for $a$ can be found in a local Euclidean
frame in which $M$ is skew-diagonal \bkop:
\eqn\defa{a= \cases{ -{1\over2} Y_{jk} \gamma^{jk} \Gamma_{11}\,
\,\,\,\,\,\,\, ({\rm for \,\,\,\, IIA}) \cr {1\over2} Y_{jk}
\sigma_3\otimes\gamma^{jk} \,\,\,\,\,\,\,\,\, ({\rm for \,\,\,\,
IIB}) }}
where $Y$ is
\eqn\defy{Y = \half Y_{i k} e^{i} \wedge e^{k} =
\sum_{r=1}^{[(p+1)/2]} \phi_{2r-1,2r} e^{2r-1} \wedge e^{2r}}
and is related to $M$ by
\eqn\fphi{M = \half M_{i k} e^i \wedge e^k= \sum_{r=1}^{[(p+1)/2]}
\tan\phi_{2r-1,2r} e^{2r-1} \wedge e^{2r}.}
In these equations, $e$ is the vielbein on the worldvolume
($g_{\mu \nu}=e^{i}{}_\mu e^{k}{}_\nu \eta_{ik}$, with $i,
k=0,\dots, p$ ). Notice that, in any orthonormal frame, if we
define the matrices $M=\sum_{i,j}M_{ij}T_{ij}$,
$Y=\sum_{i,j}Y_{ij}T_{ij}$, where $T_{ij}=e_{ij}-e_{ji}$ and
$e_{ij}$ are matrix units, then the relation between $M$ and $Y$
is \eqn\tanheq{ M=\tanh (Y).}
As a final remark about notation, we will use ${\rm
vol}_{p+1}={\sqrt {|g|}}d^{p+1}\xi$ to denote the canonical volume
element on the $p$-brane associated to the induced Riemannian
metric $g$.

\newsec{Deformed equations for BPS configurations}

In our discussion we will consider only two types of geometry:

\noindent $\bullet$ Infinite flat branes, filling a submanifold
$\IR^{p+1}\times \{pt\}\subset  \IR^{p+1} \times M_{9-p}$. These
are analyzed in section 4.

\noindent $\bullet$ Branes wrapping cycles in manifolds of
irreducible non-trivial holonomy. There is a finite number of such
cases that preserve supersymmetry and these can be summarized in a
table:

$$\vbox{\offinterlineskip\halign{ \strut # height10pt depth 13pt&
\quad#\quad\hfill\vrule&\quad$#$\quad\hfill\vrule&
\quad$#$\quad\hfill\vrule& \quad$#$\quad\hfill\vrule&
\quad$#$\quad\hfill\vrule& \quad$#$\quad\hfill\vrule\cr
\noalign{\hrule} \vrule& p+1& SU(2)&SU(3)&G_2&SU(4)&Spin(7)\cr
\noalign{\hrule} \vrule& 2& {\rm divisor/SLag}&{\rm
holomorphic}&-&{\rm holomorphic}&-\cr \noalign{\hrule} \vrule& 3&
-&{\rm SLag}&{\rm associative}&-&-\cr \noalign{\hrule} \vrule& 4&
X&{\rm divisor}&{\rm coassociative}&{\rm Cayley}&{\rm Cayley}\cr
\noalign{\hrule} \vrule& 5& -&-&-&-&-\cr \noalign{\hrule} \vrule&
6& -&X&-&{\rm divisor}&-\cr \noalign{\hrule} \vrule&7&
-&-&X&-&-\cr \noalign{\hrule} \vrule& 8& -&-&-&X&X\cr
\noalign{\hrule} }}$$
\bigskip
\centerline{Table 3. Cycles in manifolds of irreducible
non-trivial holonomy.}
\bigskip

Cycles marked in the table are those that solve $(1-\G)\eta=0$ in
the absence of $B$-field and gauge fields on the branes. In this
section we investigate how the story changes in the presence of
$B$ and gauge fields. In the cases we examine, there are no
changes in the conditions on the cycle itself.\foot{We have not
however completely ruled out the interesting possibility (raised
in \gp ) that the cycles themselves are sometimes deformed.} The
gauge fields on the other hand are found (in dimension four or
greater) to obey non-linear generalizations of the usual instanton
equations.

\subsec{SU(2) holonomy}

 We start by analyzing manifolds of $SU(2)$ holonomy. There are two
covariantly constant spinors of the same chirality, $\eta_+$ and
$\eta_-$, and we will choose them in such a way that
\eqn\spinsutwo{ \gamma_{\mb}\eta_+=0, \,\,\,\,\,\
\gamma_{mn}\eta_+=\Omega_{mn}\eta_-.} With our conventions, the
basis for spinors is given by $\eta_+$, $\gamma_{mn} \eta_+$ for
spinors of negative chirality, and $\gamma_m \eta_+$ for spinors
of positive chirality. Notice that $\gamma_{\mb \bar n}
\eta_-=-{\overline \Omega}_{\mb \bar n} \eta_+$. We have
normalized $\vert\vert \Omega \vert\vert^2=4$.

\subsubsec{$p=1$}

We first consider the case of a one-brane wrapping a 2-cycle.
Using the explicit expression for $\Gamma$ given in \gamm, the
equation $\Gamma\eta =\eta$, where $\eta$ is an ${\rm Sl}(2, \IR)$
doublet of spinors, becomes
\eqn\pone{ \eqalign{ &{{\sqrt {|g|}}\over {\sqrt{ \vert g+M
\vert}}} \bigl( 1 -
 {1 \over 2}
\gamma^{\mu \nu }M_{\mu \nu}\bigr) \Gamma_{(0)}\eta_1 =
i\eta_2,\cr &{{\sqrt {|g|}}\over {\sqrt{ \vert g+M \vert}}} \bigl(
1 +  {1 \over 2} \gamma^{\mu \nu}M_{\mu \nu}\bigr)
\Gamma_{(0)}\eta_2 =  i\eta_1.\cr }}
The second equation is actually a consequence of the first, as
follows from the identity $\Gamma^2=1$. The spinors $\eta_{1,2}$
are covariantly constant, so they must be of the form $\eta_1 =
z_+ \eta_+ + z_- \eta_-$ and $\eta_2 = w_+ \eta_+ + w_- \eta_-$,
where $z_{\pm}$, $w_{\pm}$ are constants.
 We can always normalize $\eta_1$ as
$z_{\pm}={\rm e}^{\mp i\theta}$. Using the above identities for
the $\gamma$'s, and $\gamma_{m \nb}\eta_+ = iJ _{m\nb}\eta_+$, we
find that unbroken supersymmetry requires $|w_+|^2 + |w_-|^2=2$.
We then introduce three angles $\chi$, $\phi_{\pm}$, and write
$w_+={\sqrt 2}\cos \chi {\rm e}^{i\phi_+}$ and $w_-={\sqrt 2}\sin
\chi {\rm e}^{i\phi_-}$. One finds the following equations:
\eqn\twofour{\eqalign{ f^*(\Omega) &= {i {\rm e}^{i \theta}\over
{\sqrt 2}} (\sin \chi {\rm e}^{i\phi_-}+ \cos\chi {\rm
e}^{-i\phi_+}) {{\sqrt {|g+ M|}} \over {\sqrt {|g|}}}{\rm vol}_2,
\cr f^*(J)+i M & = {{\rm e}^{i \theta}\over {\sqrt 2}} (\cos\chi
{\rm e}^{i\phi_+}-\sin \chi {\rm e}^{-i\phi_-}) {{\sqrt {|g+ M|}}
\over {\sqrt {|g|}}}{\rm vol}_2.\cr} }
The above equations can be written as \eqn\rot{ \pmatrix{f^*(J)+ i
M \cr -if^*(\Omega)} = U \pmatrix{1 \cr 0}{{\sqrt {|g+ M|}} \over
{\sqrt {|g|}}}{\rm vol }_2,} where $U$ is a constant $U(2)$ matrix . Notice
that, when $M=0$, we obtain the usual calibration condition for a 2-cycle
in a K3
manifold that the real vector $(f^*(J),{\rm Re}(f^*(\Omega)),
{\rm Im}(f^*(\Omega)))$ lies on a sphere of radius ${\rm vol}_2$
(see, for example, \hl, section V.3):
\eqn\calsutwo{
(f^*(J), {\rm Re}(f^*(\Omega)),
{\rm Im}(f^*(\Omega)))=(\cos \theta, \sin \theta \cos \phi,
\sin\theta\sin \phi) {\rm vol}_2,
}
where $\theta$, $\phi$ are constant angles along the two-cycle. The
${\bf S}^2$ that shows up in \calsutwo\ is in fact related
to the ${\bf S}^2$ of complex structures. If one chooses
a complex structure by choosing a direction, the corresponding
point on the sphere gives the holomorphic condition for the
2-cycle, while the intersection of the normal plane to this
direction with the sphere gives the ${\bf S}^1$ family of special
Lagrangian submanifolds. Of course, both are related by an $SO(3)$
rotation.

Since $M$ is an antisymmetric tensor in two dimensions, one has
$|g +M|=|g| + M^2$. Using this, it is easy to check from \rot\
that:
$$
M=\pm {{\rm Im}\,u_{11} \over \sqrt{ 1-({\rm Im}\,u_{11})^2}}{\rm vol}_2,
$$
\eqn\msutwo{
f^*(J)= \pm {{\rm Re}\,u_{11} \over \sqrt{ 1-({\rm Im}\,u_{11})^2}}{\rm
vol}_2,\,\,\,\
f^*(\Omega)={i u_{21} \over \sqrt{ 1-({\rm Im}\,u_{11})^2}}{\rm vol}_2,}
where $u_{11}$ and $u_{21}$ are the corresponding entries of the
matrix $U$, and are constant complex numbers. The equations \msutwo\ say that
the vector $(f^*(J), {\rm Re}(f^*(\Omega)),
{\rm Im}(f^*(\Omega)))$ still has the structure \calsutwo, even for a nonzero
$M$. Therefore, the Born-Infeld field $M$ does not change the
usual calibration condition.

\subsubsec{$p=3$}

For a D3-brane wrapping a four-cycle, the condition $\Gamma \eta
=\eta$ of unbroken supersymmetry reduces to
\eqn\pthree{ \eqalign{ &{{\sqrt {|g|}}\over {\sqrt{ \vert g+M
\vert}}} \bigl( 1 + {1\over 8}\gamma^{\mu \nu \rho\sigma} M_{\mu
\nu}M_{\rho\sigma}
 + {1 \over 2}
\gamma^{\mu \nu}M_{\mu \nu}\bigr) \Gamma_{(0)}\eta_1 =
-i\eta_2,\cr &{{\sqrt {|g|}}\over {\sqrt{ \vert g+M \vert}}}
\bigl( 1 + {1\over 8}\gamma^{\mu \nu \rho\sigma} M_{\mu
\nu}M_{\rho\sigma}
 - {1 \over 2}
\gamma^{\mu \nu}M_{\mu \nu}\bigr) \Gamma_{(0)}\eta_2 = i\eta_1,\cr
}}
where again $\eta = (\eta_1 \, , \eta_2)^T$. When D3 wraps the
manifold itself, when solving $(1-\G)\eta=0$ we can take into
account that, according to our conventions, $\G_{(0)}\eta=-\eta$.
Using the ansatz above for the spinors $\eta_{1,2}$, one finds
again that $|w_+|^2 + |w_-|^2=2$, and the following equations:
\eqn\twofour{ \pmatrix{{1\over 2}(J+ i M)^2 \cr  M\wedge \Omega} =
U \pmatrix{1 \cr 0}{{\sqrt {|g+ M|}} \over {\sqrt {|g|}}}{\rm vol
}_4,} where $U$ is again a $U(2)$ matrix.

We have a family of solutions depending on the value of $U$. For
example, if $U=\pmatrix {0&-1 \cr 1&0}$, then one can check that
$M={\rm Re}(\Omega)$ is a solution of the equations. To see this,
one has to use that $M^{1,1}=0$, and that $M\wedge M = 2\,{\rm
vol}_4$. If $U$ is the identity matrix, one has $M\wedge
\Omega=0$, and $(J+iM)^2/2{\sqrt {|g+ M|}}$ is a constant phase
times $d^4 \xi$. The first equation says that $M^{2,0}=0$. Using
the fact that $J^2/2$ is the volume element, one sees that
$(J+iM)^2/2{\sqrt {|g+ M|}}d^4 \xi$ is in fact a complex number of
modulus one. For $U={\bf 1}$, the above equations can then be
written as: \eqn\twofourag{\eqalign{& J\wedge M= k \, ({\rm vol}_4
-{1 \over 2} M\wedge M), \cr &M^{2,0}=0, \cr}} where $k$ is a
constant.

For a compact 4-cycle $\Sigma^4$ the value of $k$ in \twofourag\
is determined in terms of the topological data. Let the closed two
forms $C_I$, $I=1,2,...b_2$ be an integer basis for
$H^2(\Sigma^4,R)$, and $I_{IJ}=\int C_I \wedge C_J$ the
corresponding intersection matrix. Then in cohomology we can
expand \eqn\trjk{\eqalign{F&=F^IC_I,\,\,\,\,\,\ B=B^IC_I,\cr
M&=M^IC_I,\,\,\,\,\,\ J=J^IC_I .\cr}} We note that $2\pi F^I$ are
integrally quantized, but  $B^I$ and $J^I$ are not quantized. From
the definition of $M$, $M^I=2\pi\apm (F^I+B^I)$. Integrating
\twofourag\ then yields
\eqn\keqs{(I_{KL}J^KJ^L-I_{KL}M^KM^L)k  =2I_{IJ}M^IJ^J .}
If $M$ and $J$ are orthogonal then either they must have the same norm
and $k$ is undetermined, or $k$ must be zero.

Another way of phrasing the conditions \twofourag\ is that $J+ k
M$ is a closed $(1,1)$-form such that
\eqn\mongeamp{ (J+ k M ) \wedge (J+ k M) = (1+ k^2) J \wedge J }
If we can write $J + k M = T + i \p \bar \p \phi$, with $\phi$ a
globally well-defined real scalar and $T$ a positive $(1,1)$ form
in the cohomology class $[J] + k [M]$ then \mongeamp\ is just the
Monge-Ampere equation for $\phi$, and there is a unique solution
\yauremark.

 Yet another   form of the conditions \twofourag\ for
supersymmetry can be obtained by decomposing $M=M^-+M^+$ into
selfdual and antiselfdual parts. The second equation in
\twofourag\ then implies $M^+=\phi J$ for some scalar $\phi$. The
first equation can be solved for $\phi$ as a function of $M^-$ and
$k$ (assuming $k\not=0$):
\eqn\trp{\phi=-{1 \over k}\bigl(1 \pm\sqrt{1+k^2(1+\vert\vert
M^-\vert \vert^2)}\bigr),}
 where $\half M^- \wedge M^{-}=-
\vert\vert M^- \vert\vert^2 {\rm vol}_4.$ The three components of
$M^-$ are then constrained by the condition $dM=0$, which becomes
\eqn\rtl{dM^-=\mp{k\over 2}  J \wedge d\vert\vert M^-\vert \vert^2
{ 1 \over \sqrt{1+k^2(1+\vert\vert M^-\vert \vert^2)}} .}

\subsec{SU(3) holonomy}

To analyze the conditions for unbroken supersymmetry, we follow
the conventions for covariantly constant spinors of \bbs.\foot{In the 
following we assume that the Calabi-Yau is compact and has
generic $SU(3)$ holonomy.}
 There
are two covariantly constant spinors $\eta_{\pm}$ of opposite
chirality, and conjugate to each other: $\eta_-^*=\eta_+$. They
are chosen in such a way that:
\eqn\gology{\gamma_{\mb} \eta_+ = \gamma_m \eta_- =0,\,\,\,\,\
\gamma_{mnp}\eta_+ = \Omega_{mnp}\eta_- , }
where $m$ is a holomorphic coordinate index, $m=1,2,3$.
 The spinor space is spanned by $\eta_+$, $\eta_-$,
$\gamma_m \eta_+$ and $\gamma_{\mb}\eta_-$. We will also need the
following identities:
\eqn\gologyii{\eqalign{&\gamma_{{\bar q} m n p} \eta_+ = {1 \over
2}\Omega_{mnp} \gamma_{{\bar q}}\eta_-, \cr &\gamma_{{\bar q} \mb
n p}\eta_+ = (g_{n \mb}g_{p{\bar q}}
 -g_{n{\bar q}}g_{p\mb})\eta_+,\cr
&\gamma_{mn}\eta_+={1\over 2} \Omega_{mnp}g^{p {\bar q}}
\gamma_{\bar q} \eta_-.\cr} }

We now analyze the conditions for unbroken supersymmetry for the
different cycles of dimension $p+1$.

\subsubsec{$p=1$}

 For a one-brane wrapping a two-cycle we can use
again \pone. Setting, as in the previous subsection, $\eta_1 = z_+
\eta_+ + z_- \eta_-$ and $\eta_2 = w_+ \eta_+ + w_- \eta_-$, one
finds that $(z_+/w_+)=-{\rm e}^{i\theta}$ is a phase, and the
equations read
 \eqnn\twosix
 $$ \eqalignno{ &f^*(J) + iM= e^{i\theta}
{{\sqrt {|g +M|}} \over {\sqrt{ \vert g \vert}}} \, {\rm vol}_2,
&\twosix (a)\cr &dX^m \wedge dX^n\Omega_{mnp}=0, &\twosix (b) \cr}
$$ where $d$ denotes the exterior derivative on the worldvolume.
In the above equation, and in similar equations in this section,
the $X^m$ denote the coordinates of the embedding.

Equation \twosix (b) implies that the cycle is holomorphically
embedded. A quick way to see this is to use local complex
coordinates in the static gauge, and normalize $\Omega_{123}=1$.
If we denote by $X^m$ the complex coordinates for the threefold,
the embedding will be described by two functions $X^2=X^2(X^1,
X^{\bar 1})$, $X^3=X^3(X^1, X^{\bar 1})$, where we have identified
$X^1$, $X^{\bar 1}$ with the complex coordinates on the one-brane
worldvolume. The second equation in \twosix\ says that
$\partial_{\bar 1}X^2=\partial_{\bar 1}X^3=0$, and the embedding
is holomorphic. Therefore, the two-cycles in threefolds are still
holomorphic. This implies that $f^*(J)={\rm vol}_2$, and the other
equation for BPS configurations (the first equation in \twosix)
says that $M$ is a constant multiple of the volume form:
\eqn\defrmpone{M= 2\pi \alpha' (F + f^*(B)) = \tan \theta \,
\vol_2.
 }
If we fix the topology of the Chan-Paton line bundle:
$\int_{\Sigma_2} F = 2 \pi n$, and the background field $B$, the
constant $\tan \theta$ is completely determined by integrating the
equation \defrmpone:
\eqn\constant{ 2\pi n + \int_{\Sigma_2} f^*(B)  = {\tan\theta
\over 2\pi \alpha'} \, \int_{\Sigma_2}  J .}
Equation
\defrmpone\
represents the only deformation of the usual equations in the
presence of $B$, for $p=1$ in a threefold. The content is simply
that ${F \over 2 \pi}$ is any integral harmonic form on any
holomorphic cycle.

\subsubsec{$p=2$}

Let's now consider the case of $p=2$ ({\it i.e.} a D2 brane
wrapping a $3$-cycle in IIA theory). Using the explicit expression
\gamm, we find that the deformed supersymmetry equation is:
\eqn\defrmi{ -{i\over {\sqrt {|g+M|}}} {1\over 3!} \epsilon^{\mu
\nu \rho}\biggl(\gamma_{\mu \nu \rho} + 3 M_{\mu \nu } \gamma_\rho
\Gamma_{11} \biggr) \eta  =\eta }
where $\eta = z_+\eta_+ +  z_- \eta_-$. We find again that
$(z_-/z_+)=-i{\rm e}^{i \theta}$ is a phase, and the equations
read:
\eqnn\threesix $$ \eqalignno{&f^* (\Omega)  = e^{i \theta}
{\sqrt{|g + M| } \over \sqrt{|g|} } \vol_3 , &\threesix (a)\cr
&f^*(J) + iM = 0 .&\threesix (b)\cr} $$
Since $f^*(J)$ is a real differential form, it follows from the
second equation that $M=0$, and one recovers the special
Lagrangian condition of \bbs.  Hence the possibility of gauge
field strengths does not lead to new BPS configurations.

\subsubsec{$p=3$}

For D3 on a four-cycle, using \pthree, we find that unbroken
supersymmetry requires $(z_+/w_+)^* = (z_-/w_-) = -i e^{i\theta},$
where $\theta$ is a constant and
\eqnn\foursix $$ \eqalignno{ &\half (f^*(J) + iM)\wedge(f^*(J) +
iM) = e^{i\theta} {{\sqrt {|g +M|}} \over {\sqrt{ \vert g \vert}}}
\, {\rm vol}_4,&\foursix (a)\cr &f^*(\Omega) \wedge dX^{\bar q}
g_{{\bar q} r} + M \wedge dX^m\wedge dX^n \Omega_{mnr}=0.&\foursix
(b) \cr}$$
We can see that when $M=0,$ one has $\theta=0$ as well (just by
reality of \foursix) and we quickly recover the original condition
of \bbs\ that the four-cycle is holomorphic.

In fact even when $M$ is nonzero the four-cycle is holomorphic. As
before, we can do the analysis in local complex coordinates. We
will assume that the embedding can be described
$X^3=X^3(X^1,X^{\bar 1}, X^2, X^{\bar 2})$. At any given point we
can always choose a frame in which the metric has the standard
form $g=(1/2)\sum_{i=1}^3 (dX^i \otimes dX^{\bar \i}+ dX^{\bar \i}
\otimes dX^i)$, and $\Omega=dX^1\wedge dX^2 \wedge dX^3$. As our
equations only involve first derivatives, we can work pointwise.
In this coordinate system, the second equation of \foursix\ can be
written as follows: \eqn\nodefone{ \eqalign{ &\alpha_{\bar 1}M_{1
\bar 2} -\alpha_{\bar 2} M_{1 \bar 1} -\alpha_1 M_{\bar 1 \bar 2}
={1 \over 2} \alpha_{\bar 2},\cr &\alpha_{\bar 2}M_{2 \bar 1}
-\alpha_{\bar 1} M_{2 \bar 2} -\alpha_2 M_{\bar 2 \bar 1} ={1
\over 2} \alpha_{\bar 1},\cr &M_{\bar 1 \bar 2} = {1 \over 2}
({\overline {\alpha_1}}\alpha_{\bar 2} -{\overline \alpha_2}
\alpha_{\bar 1 }),\cr}} where $\alpha_i=\partial_i X^3$,
$\alpha_{\bar \i}=\partial_{\bar \i} X^3$, $i=1,2$. There is,
however, an extra constraint that one has to fulfill: the $2$-form
$M$ is a real form, and in particular it satisfies $M_{1 \bar
1}^*=-M_{1 \bar 1}$, $M_{2 \bar 2}^* = -M_{2\bar 2}$. If we write
these reality conditions using the explicit expressions in
\nodefone, we find that the following equation has to be
satisfied: \eqn\nodeftwo{ |\alpha_{\bar 1}|^2+|\alpha_{\bar 2}|^2
+ |\alpha_1|^2|\alpha_{\bar 2}|^2 + |\alpha_2|^2|\alpha_{\bar
1}|^2 - 2 {\rm Re} (\alpha_1 \alpha_{\bar 1} {\overline \alpha_2}
{\overline \alpha_{\bar 2}}) =0.} The sum of the last three terms
is greater than or equal to $(|\alpha_1||\alpha_{\bar
2}|-|\alpha_2||\alpha_{\bar 1}|)^2$, therefore positive, and it
follows then that $\alpha_{\bar 1}=\alpha_{\bar 2}=0$. The
embedding is holomorphic, and the four-cycle has to be a divisor.

We can now look at the equations for the gauge field $M$. From the
last equation in \nodefone\ it follows that $M^{2,0}=0$. If the
cycle is holomorphically embedded, \foursix\ gives an equation for
the $(1,1)$ part of $M$:
\eqn\swfour{ f^*(J)\wedge M =\tan \theta ({\rm vol}_4 -{1 \over 2}
M \wedge M).}
These equations are non-linear deformations of the usual instanton
equations $M^{2,0}=0$, $f^*(J)\wedge M = k {\rm vol}_4$, where
$f^*(J)$ is the K\"ahler form on the four-cycle. The BPS
configuration we have found is then a divisor in a Calabi-Yau
threefold together with a deformed instanton on it. Notice that the
above equations are precisely the equations \twofourag\ that one
finds for a D3 brane wrapping a manifold of $SU(2)$ holonomy. Thus,
by the discussion surrounding \mongeamp\ there is a unique
solution, as long as the cohomology class $[f^*(J) + k M]$ is in
the K\"ahler cone of the 4-cycle.

\subsubsec{$p=5$}

If the fivebrane wraps the six-cycle itself, we have
\eqn\pfive{ \eqalign{ &{{\sqrt {|g|}}\over {\sqrt{ \vert g+M
\vert}}} \bigl( 1 + {1\over 8}\gamma^{\mu \nu \rho \sigma} M_{\mu
\nu } M_{\rho \sigma}
 + {1 \over 2} \gamma^{\mu \nu }M_{\mu \nu} +
{1\over 48} \gamma^{\mu \nu\rho \sigma \tau \upsilon} M_{\mu
\nu}M_{\rho \sigma}M_{\tau \upsilon}\bigr) \Gamma_{(0)}\eta_1 =
i\eta_2,\cr &{{\sqrt {|g|}}\over {\sqrt{ \vert g+M \vert}}} \bigl(
1 + {1\over 8}\gamma^{\mu \nu\rho \sigma} M_{\mu \nu}M_{\rho
\sigma}
 - {1 \over 2}
\gamma^{\mu \nu}M_{\mu \nu} -{1\over 48} \gamma^{\mu \nu\rho
\sigma\tau \upsilon} M_{\mu \nu}M_{\rho \sigma}M_{\tau
\upsilon}\bigr)
 \Gamma_{(0)}\eta_2 =  i\eta_1.\cr }}
These equations imply that:
\eqnn\sixsix $$ \eqalignno{ &{1 \over 3!} (J + iM)^3 = e^{i\theta}
{{\sqrt {|g +M|}} \over {\sqrt{ \vert g \vert}}} \, {\rm vol}_6, &
\sixsix (a) \cr &(g^{{\bar q} p} -M^{{\bar q} p})
\Omega_{pmn}M^{mn}=0. & \sixsix (b) \cr}$$
The last equation can be analyzed as follows. Define
$v_p=\Omega_{pmn}M^{mn}$, and consider $v_q^{*}(g^{{\bar q} p}
-M^{{\bar q} p})v_p$, which is zero by \sixsix (b). Using that $M$
is real and antisymmetric, one finds that $v_q^{*}g^{{\bar q} p}
v_p=0$, hence $\Omega_{pmn}M^{mn}=0$. But this means that
$M^{2,0}=0$. We can then write the equations \sixsix\ as
\eqn\sisixbis{ \eqalign{ &{1 \over 2!}J\wedge J\wedge M - {1 \over
3!} M \wedge M \wedge M  = \tan\theta ({\rm vol}_6 - {1\over 2!}J
\wedge M\wedge M), \cr &M^{2,0}=0.}} As in \keqs\ the value of the
constant $\theta$ can be determined in terms of the topological
data by integration over the six-cycle.

\subsec{$G_2$ holonomy}

To analyze the supersymmetry conditions in manifolds of $G_2$
holonomy, we need some facts about spinors in such manifolds. We
will identify the spinors
 with the octonions $\IO$ in the
Cayley-Dickson description: an octonion will be given by a pair of
quaternions $(a,b)$, where $a=x^8+ x^1i + x^2j+x^3 k$, and $b=x^4
+x^5i + x^6j+x^7 k$. The multiplication rule is $(a,b) \cdot
(c,d)=(ac-{\overline d} b, da + b {\overline c})$, where the
overline denotes the usual quaternion conjugation. An octonion
written in this way is imaginary if $a$ is an imaginary quaternion
({\it i.e.} if $x^8=0$ in the above description). The seven
imaginary units are then $(i,0), \dots, (0,k)$. We can identify
$\IR^7 \simeq {\rm Im}\, \IO$ in the obvious way, and Clifford
multiplication is therefore given by octonionic multiplication by
the imaginary units. The $\gamma$ matrices will be $i$ times the
imaginary units acting through multiplication, in order to have
$\{ \gamma_\mu, \gamma_\nu \}=2\delta_{\mu \nu}$ in flat space.
They are then $8 \times 8$ imaginary, antisymmetric matrices.

In a manifold of $G_2$ holonomy there is a covariantly constant
spinor $\vartheta$ which in the above representation can be taken
as $(1,0)$ ({\it i.e.} the unit octonion). A basis for the spinor
space is then given by $\vartheta$, $\gamma_\mu \vartheta$,
$\mu=1, \dots, 7$. We also have a calibration $\Phi$, which is a
closed three-form, and the following identities \kmt: \eqn\ident{
\gamma_{\mu \nu \rho} \vartheta =i \Phi_{\mu \nu \rho}\vartheta - (*\Phi)_{\mu \nu \rho \lambda} \gamma^{\lambda} \vartheta ,
\,\,\,\,\,\ \gamma_{\mu \nu }\vartheta =\Phi_{\mu \nu
\rho}\gamma^\rho \vartheta, \,\,\,\,\,\ \gamma_{\mu \nu \rho
\sigma}\vartheta =(*\Phi)_{\mu \nu \rho \sigma}\vartheta - 4i \Phi_{[\mu \nu \rho} \gamma_{\sigma ]}\vartheta .}

We can already analyze the conditions for unbroken supersymmetry
in the presence of Born-Infeld fields. For a D2 brane, we get the
equations: \eqnn\gthree $$ \eqalignno{ &f^*(\Phi) = {{\sqrt {|g
+M|}} \over {\sqrt{ \vert g \vert}}} \, {\rm vol}_3, &\gthree
(a)\cr &M\wedge dX^\mu=0.&\gthree (b)\cr} $$ The second equation
implies
 $M=0$. Hence we recover
the usual condition for an associative cycle in a $G_2$ manifold.

The analysis for a D3 brane is similar, and we obtain the
following conditions: \eqnn\gfour $$ \eqalignno{
&f^*(*\Phi)-{1\over 2}M\wedge M = {{\sqrt {|g +M|}} \over {\sqrt{
\vert g \vert}}} \, {\rm vol}_4, &\gfour (a)\cr &M\wedge  dX^\mu
\wedge dX^\nu \Phi_{\mu \nu \rho}=0.&\gfour (b)\cr} $$ The
equation \gfour (b) can be interpreted as follows. The ${\bf 21}$
of ${\rm Spin}(7)$ decomposes under $G_2$ as ${\bf 21}={\bf 14} +
{\bf 7}$. \gfour (b) says that the 2-form $M\wedge dX^\mu \wedge
dX^\nu$ belongs to the ${\bf 14}$. Using the projector of the
${\bf 21}$ of ${\rm Spin}(7)$ onto the ${\bf 7}$ \figue
\eqn\projg{ {\bf P}_{\mu \nu \rho \sigma}={1 \over 6} (g_{\mu
\rho}g_{\nu \sigma} - g_{\mu \sigma} g_{\nu\rho} -(*\Phi)_{\mu \nu
\rho \sigma}),} we can write \gfour (b) as
\eqn\gfouralt{ {\bf P}(M\wedge  dX^\mu  \wedge dX^\nu)=0.}

In the case when the 4-cycle is a coassociative 4-fold the
equation \gfouralt\ implies that $M$ is an anti-self-dual 2-form
on the D3 brane worldvolume: $M^+=0$. This can be easily proved by
working in local coordinates. Another proof proceeds as follows.
If the cycle is coassociative then we may replace $f^*(*\Phi)
={\rm vol}_4$, and divide through by ${\rm vol}_4$. We then square
the equation and use
\eqn\matrxd{ {|g +M| \over   \vert g \vert} = 1 - \half {\Tr} M^2
+ \det M }
(here $M$ is an antisymmetric matrix, in local coordinates). Then
the square of \gfour(a) becomes $(1- {\rm Pf}(M))^2= 1- \half
{\Tr}(M^2) + \det M$ so ${\Tr} (M^+)^2 = 0 $.

 When a D6 brane wraps a $G_2$ manifold, the conditions for
unbroken supersymmetry give equations for the gauge field. These
will be analyzed in section 4 using the group-theory approach.

\subsec{SU(4) holonomy} On a manifold of $SU(4)$ holonomy there
are two covariantly constant spinors, $\eta_{\pm}$, with the same
chirality and complex conjugate to each other. They are chosen in
such a way that $\gamma_\mb \eta_+=0$. We have the following
identities, \eqn\identfour{
\gamma_{mnpq}\eta_+=\Omega_{mnpq}\eta_-, \,\,\,\,\ \gamma_{{\bar
q} mnp}\eta_+ =3i J_{{\bar q} [ m} \gamma_{np]}\eta_+. } The
positive chirality spinor space is spanned by $\eta_+$,
$\gamma_{mn}\eta_+$, and $\eta_-$, while the negative chirality
spinor space is spanned by $\gamma_m \eta_+$,
$\gamma_{mnp}\eta_+$. Supersymmetric cycles in fourfolds were
recently considered in \gkv.

\subsubsec{$p=1$} For a D1 brane wrapping a 2-cycle in a
Calabi-Yau fourfold, we find two equations. The first one is
\twosix (a), while the second one is \eqn\twoeight{ d X^m \wedge d
X^n =0.} This implies that the cycle is holomorphic. $M$ is
constrained in an analogous way to the $SU(3)$ case.

\subsubsec{$p=3$} For a D3 brane wrapping a four-cycle, the
conditions we obtain are more complicated. Set $z_{\pm}={\rm
e}^{\mp i\theta/2}$. The analysis of the equations gives
$w_+=i{\rm e}^{i \phi/2}$, $w_-=i{\rm e}^{-i\phi/2}$, and the
following conditions:
\eqnn\foureight $$ \eqalignno{ &-{1\over 2}(f^*(J)+ iM)^2 +   f^*(
\overline \Omega_\theta) ={\rm e}^{i(\phi+\theta)/2}{{\sqrt {|g
+M|}} \over {\sqrt{ \vert g \vert}}} \, {\rm vol}_4,&\foureight
(a) \cr &{\rm Im}({\rm e}^{i(\phi+\theta)/2}f^*(\Omega_\theta
))=0,&\foureight (b)\cr &(f^*(J) + i M)\wedge \bigl( d X^n \wedge
d X^p +{1 \over 2} ({\overline \Omega_\theta})_{{\bar p}{\bar
q}}{}^{np} d X^{\bar p}\wedge d X^{\bar q} \bigr)=0.&\foureight
(c)\cr} $$
where $\Omega_\theta := {\rm e}^{-i\theta}\Omega$. To write the
last equation, we have used that $\gamma_{\mb {\bar n}}
\eta_{-}=-{1 \over 2}
 {\overline \Omega}_{{\bar m}{\bar n}}{}^{pq}\gamma_{pq}\eta_+$, with
 the normalization $\vert\vert \Omega \vert \vert ^2=16$. Again,
notice that when $M=0$ the reality of $f^*(J)$ imposes that
$\theta+ \phi=0$. The second equation gives ${\rm Im}(
\Omega_\theta)=0$ and the first equation reads, \eqn\caylcal{
-{1\over 2}f^*(J)^2 + {\rm Re}(   f^*(\Omega_\theta)) ={\rm
vol}_4,} which is the usual condition for a Cayley calibration
\hl\ obtained in this context in \bbcia.

\subsubsec{$p=5$}

Finally, we can analyze the deformed equations for a six-cycle in
a fourfold. If we set $z_{\pm}={\rm e}^{\mp i\theta/2}$, we find
again $w_+=i{\rm e}^{i \phi/2}$, $w_-=i{\rm e}^{-i\phi/2}$, and
three equations which are similar to \foureight: \eqnn\sixeight $$
\eqalignno{ &{1\over 3!}(f^*(J)+ iM)^3 -i M\wedge f^*( \overline
\Omega_\theta) =-i{\rm e}^{i(\phi+\theta)/2}{{\sqrt {|g +M|}}
\over {\sqrt{ \vert g \vert}}} \, {\rm vol}_6,&\sixeight (a) \cr
&{\rm Im}(M\wedge {\rm
e}^{i(\phi+\theta)/2}f^*(\Omega_\theta))=0,&\sixeight (b)\cr
&(f^*(J) + i M)^2\wedge \bigl( d X^n \wedge d X^p +{1 \over 2}
({\overline \Omega_\theta})_{{\bar p}{\bar q}}{}^{np} d X^{\bar
p}\wedge d X^{\bar q} \bigr)=0.&\sixeight (c)\cr} $$
where again $\Omega_\theta := {\rm e}^{-i\theta}\Omega$. Notice
that when $M=0$ one recovers the usual conditions for a
holomorphic embedding ({\it i.e.}, the cycle is a divisor).

For a D7 brane wrapping a $SU(4)$ manifold, the analysis of the
equations along these lines is more involved. As it will become
clear in section 4, one finds a natural generalization of
\sisixbis\ that can be obtained much more easily using the
group-theory approach.

\subsec{Spin(7) holonomy}

To analyze the unbroken supersymmetries in manifolds of ${\rm
Spin}(7)$ holonomy, we first set the relevant spinor algebra. We
will regard the spinors of positive or negative chirality as
octonions: $S^+ \simeq S^- \simeq \IO$. The Clifford algebra
$C\ell_8$ is represented by \eqn\eightdeuc{ \eqalign{
\Sigma^{1,\dots, 8} & = \pmatrix{ 0 & -i\gamma^{1,\dots, 8}\cr
-i\tilde \gamma^{1,\dots, 8} & 0 \cr} \cr \tilde \gamma^{1,\dots,
7 } & = - \gamma^{1,\dots, 7 }\cr \tilde \gamma^{8 } & = \gamma^{8
}\cr} } where $\gamma^i$ is the representation of $C\ell_7$
described above, and $\tilde \gamma^{1,\dots, 7}$ is  the other
inequivalent representation of $C\ell_7$ (notice that $-i
\gamma^{1, \dots, 7}$ is given by octonionic multiplication by the
imaginary units).  We take
 $\gamma^8_{a \dot a} = i\delta_{a \dot a}$. The chirality operator is
\eqn\chiral{ \bar \Sigma   = \pmatrix{ -1_8 & 0 \cr 0 & 1_8\cr} }
Note that $(\Sigma^i)^{T} = \Sigma^i$, and $ (\Sigma^i)^{*} =
\Sigma^i$, $i=1,\dots, 8$.

We choose the embedding of ${\rm Spin}(7)$ in ${\rm Spin}(8)$ of
\figue, in which the spinor representation decomposes as ${\bf
8}_s \rightarrow {\bf 1} + {\bf 7}$.

In a manifold of ${\rm Spin}(7)$ holonomy there is one covariantly
constant spinor, which we will take of positive chirality. We will
denote it by $\vartheta$ again, and using the identification
$S^+\simeq \IO$ this spinor can be regarded as the unit octonion.
We also have a calibration $\Omega$, which in this case is a
closed four-form. In terms of the calibration on manifolds of
$G_2$ holonomy, we have (in an orthonormal basis): \eqn\calib{
\Omega=\Phi \wedge dx^8 + *\Phi.} Notice that this differs from
the conventions in \hl. The calibration satisfies
$*\Omega=\Omega$.

The positive chirality spinor space is spanned by $\vartheta$ and
$\Sigma_{\mu \nu }\vartheta$, $\mu,\nu=1, \dots, 8$. It is
important to notice that the $\Sigma_{\mu \nu}$ are not
independent: they obey the self-duality condition \eqn\selfdual{
\Sigma_{\mu \nu} \vartheta =-{1 \over 6} \Omega_{\mu \nu}{}^{\rho
\sigma}\Sigma_{\rho \sigma} \vartheta.} This means that the tensor
$\Sigma_{\mu \nu}\vartheta$ in the  ${\bf 28}$ of ${\rm Spin}(8)$
belongs to the ${\bf 7}$ of ${\rm Spin}(7)$. Therefore, only seven
components are linearly independent and we find the right counting
of generators for $S^+$. The generators of $S^-$ are simply given
by $\Sigma_\mu\vartheta$, $\mu=1, \dots, 8$. To analyze the
unbroken supersymmetries we will need the following identity \kb\
\eqn\ident{ \Sigma_{\mu \nu\rho \sigma}\vartheta =\Omega_{\mu
\nu\rho \sigma}\vartheta - 
\Omega^{\lambda}_{\,\,[\mu\nu\rho} \Sigma_{\lambda]\sigma} \vartheta,} and the expression for the projector
${\bf 28} \rightarrow {\bf 7}$ \figue: \eqn\projsps{ {\bf P}_{\mu
\nu\rho \sigma}={1 \over 8}(g_{\mu \rho}g_{\nu \sigma} - g_{\mu
\sigma} g_{\nu\rho} -\Omega_{\mu \nu\rho \sigma}).}

We can now analyze a D3 brane wrapping a four-cycle in a manifold
of ${\rm Spin}(7)$ holonomy. After some straightforward algebra,
one finds
\eqnn\cayley $$ \eqalignno{ &f^*(\Omega) -{1\over 2} M\wedge M
={{\sqrt {|g +M|}} \over {\sqrt{ \vert g \vert}}} \, {\rm vol}_4,
&\cayley (a)\cr & {\bf P}(M\wedge dX^\mu \wedge d X^\nu)=0.
&\cayley (b)\cr} $$ Notice that, for $M=0$, we recover the fact a
supersymmetric cycle is Cayley \bbcia.

In a manner similar to the case of $G_2$ holonomy,  in the case
when the 4-cycle is a Cayley 4-fold the equation \cayley{b}\
implies that $M$ is an anti- self-dual 2-form on the D3 brane
worldvolume: $M^+=0$. Again, this  can be easily proved by working
in local coordinates, or using exactly the same argument as in the
$G_2$ case by squaring  \cayley{a}.

 Again, the case of the D7 brane wrapping a
${\rm Spin}(7)$ manifold is more involved using these techniques,
and will be considered in section 4.

\subsec{A comment on the equations}

To write the above conditions for deformed cycles, we have
decomposed $\Gamma \eta =\eta$ in the appropriate basis of the
spinor space and we have written the equations that one derives
for the different independent elements involved in the equations.
For example, \twosix (a) gives the piece proportional to $\eta_+$,
while \twosix (b) give the piece proportional to $\gamma^p\eta_+$.
However, as it has been pointed out in \bbs\bbcia, to find
solutions of this equation it is enough to solve the equation
$\eta^{\dagger} \Gamma \eta =\eta^\dagger\eta$, which usually
gives only one equation (as the inner product with
$\eta^{\dagger}$ kills the components which are orthogonal to
$\eta$). The reason that these two procedures are equivalent is
the following. If we denote $P_-={1\over 2}(1-\Gamma)$, we see
from the properties of $\G$ that $P_-$ is an Hermitian projector,
$P_-^{\dagger}=P_-$, $P_-^2= P_-$. The condition for unbroken
supersymmetry can be written as $P_-\eta=0$, but this is
equivalent to
\eqn\integr{  \bigl(P_-\eta)^{\dagger}P_-\eta=
\eta^{\dagger}P_-\eta =0.}
Therefore, solving $P_-\eta=0$ is equivalent to solving
$\eta^{\dagger}P_-\eta=0$. This implies, in particular, that the
additional equations in $P_-\eta=0$ are in fact consequences of
$\eta^{\dagger}P_-\eta=0$. This last equation gives the conditions
labeled as (a) in this section. For $M=0$, they give the standard
definitions of calibrations. The fact that the other equations
(labeled as (b) and (c)) follow from this one is not obvious from
a mathematical point of view. In the $M=0$ case, they give
additional properties of the calibrated submanifolds. For example,
for $p=1$ \twosix (b) is equivalent to holomorphicity, which in
turn is implied by \twosix (a) when $M=0$. On the other hand,
these additional equations can show features which are not
manifest in the main equations (a). This is one of the reasons
that we have decided to spell them out in detail. As we will see
in the next section, they can be extremely useful once the $M$
field is included. The mathematical meaning of equations (b), (c),
at least for the standard calibrations, is the following: the
condition that a submanifold is calibrated can be stated in terms
of a differential system ${\psi_j}$, $j=1, \dots, n$, where the
$\psi_j$ are differential forms on the ambient space \hl. A
submanifold $W$ is calibrated if and only if the forms $\psi_j$
restrict to zero on $W$. The equations (b), (c) that we have found
are in fact part of the system of equations associated to this
differential system.

As a final remark, notice the appearance in the deformed equations
of the complexified K\"ahler form, since $f^*(J) + iM =f^*(J +
2\pi i \alpha' B)+ 2\pi i \alpha' F$.

\subsec{Instanton Actions}

One application of this work is to further study of mirror
symmetry. In particular, in \bbs\   mirror symmetric formulae for
sums over D-brane and M-brane instantons were suggested. It
remains a challenging problem to make these formulae concrete and
test them. One important ingredient in the sums over D-brane
instanton corrections are the instanton actions. It is worth
noting that the real part of the instanton actions can easily be
derived from the above equations. \foot{The imaginary part, by
contrast, is a much more subtle quantity, and is discussed in
\hm\wittendinst.} As an illustration consider the D-instanton
effects in IIB string theory on a CY 3-fold. We must consider
$p=-1,1,3,5$. The case $p=-1$ has not been discussed since it
doesn't lead to interesting worldvolume equations. For the case
$p=1$ wrapping a 2-cycle $W_2$ we integrate equation \twosix{a}\
and then substitute the result \constant\ to get:
\eqn\instacti{ Re(I) = -{T_1 } \sqrt{\biggl(\int_{W_2} J\biggr)^2
+ \biggl( \int_{W_2} M\biggr)^2 } }
as expected from the tension formula for the $SL(2,\IZ)$ multiplet
of strings. Similarly, for a D3 wrapping a 4-cycle $W_4$ we get,
in a similar way
\eqn\instactii{ Re(I) = -{T_3 } \sqrt{\biggl(\int_{W_4} J\wedge M
\biggr)^2 + \biggl( \int_{W_4} \half J \wedge J - \half M \wedge M
\biggr)^2 } }
in accord with the 2-brane and 0-brane charges induced by the
Chan-Paton bundle. Finally, for a D5 wrapping the full Calabi-Yau
3-fold
\eqn\instactiii{ Re(I) = -{T_5 } \sqrt{\biggl(\int_{W_6} {1\over
3!} J^3 - {1\over 2!} J M^2 \biggr)^2 + \biggl( \int_{W_6} \half
J^2 M  - {1\over 3!}  M^3  \biggr)^2 } }
Again, this is in accord with the standard formulae for induced
D-brane charges from the Chan-Paton bundle, to leading order in
$\alpha'$.

\newsec{Group-theoretical basis for the deformed instanton equations }

Since we have found that the cycles are not deformed, we now
consider Euclidean flat branes wrapping a submanifold $\IR^{p+1}
\times \{ pt \} \subset \IR^{p+1} \times M_{9-p}$ and the
deformation of the instanton equation on them. We will exploit
here the fact that the $\Gamma$ matrix can be written in the
rotated form \gammaaa. We want to solve the equation \proj, where
$\eta$, in the type IIB theory, is a doublet of spinors $\eta_i$,
$i=1,2$, and $\G$ depends on the Born-Infeld field $M$. Suppose
that we find a covariantly constant spinor $\chi$ in an
irreducible representation of ${\rm Spin}(p+1)$ satisfying the
equation \eqn\aneq{ \half Y_{ij}\gamma^{ij}\chi=k\chi,} where $k$
is a constant scalar. Then, the equation \proj\ is easily solved
by setting $\eta_1=\chi$, $\eta_2 =\pm i ^{(p+3)/2}{\rm
e}^{k}\chi$ (where the sign depends on the chirality of $\chi$).

The equation \aneq\ is a simple equation for unbroken
supersymmetry in terms of the field $Y$. However, since there is a
nonlinear relation  between $Y$ and $M$ given by \tanheq, the
conditions for unbroken supersymmetry of $M$ can be complicated.
$M$ on the other hand obeys the simple relation $dM=0$, which is a
complicated constraint on $Y$.

Some understanding of the relation \aneq\ follows simply from
group theory. Let us regard the antisymmetric matrix $Y_{ij}$ (in
a local orthonormal frame) as an element of the Lie algebra ${\rm
spin}(p+1)$.  The equation \aneq\ simply says that the
infinitesimal rotation by $Y$ preserves $\chi$ up to a rescaling.
Let $h_\parallel \subset {\rm spin}(n)$ be the Lie subalgebra
stabilizing the one-dimensional space spanned by $\chi$. Thus
elements of $h_\parallel$ act on $\chi$ by a (possibly vanishing)
constant. Let    $h_\perp$ be the orthogonal complement of $h$ in
the Killing metric. Elements in $h_\perp$ rotate $\chi$ to a
nonzero orthogonal  spinor. Then
\eqn\decliealg{{\rm  spin}(n) = h_\parallel \oplus  h_\perp . }
The equations \aneq\ simply say that $Y_\perp=0$, and $Y_\parallel
= {\rm constant}$, in an obvious notation. Now these translate
into conditions on $(\tanh Y)_\parallel$ and $(\tanh Y)_\perp$.
The relation between $(\tanh Y)_\parallel $ and $(\tanh Y)_\perp$
is complicated in general, although it is constrained by group
theory.

In the following  subsections we will analyze the condition \aneq\
and the resulting equations for $M$, in various dimensions, using
this group-theoretic approach.

\subsec{$p=3$ case and deformed instanton equation in four
dimensions}

We begin with the case of $p=3$ in four-dimensional Euclidean
space with metric $g_{\mu\nu}=\e \delta_{\mu\nu}$. The Lie algebra
$so(4)$ is the representation $(3,1) \oplus (1,3)$ of $su(2)
\oplus su(2)$. The choice of spinor singles out the stabilizer
subgroup $h= (u(1),0)$.   Let $Y$ be a $4\times 4$ antisymmetric
real matrix. Define $Z:=\tanh Y$ by the power series $Z =
\sum_{m=0}^\infty a_m Y^{2m +1} $. ($a_m$ can be written in terms
of Bernoulli numbers, but we will not need this.) Note that $Z$ is
antisymmetric and real. For an antisymmetric matrix, we denote by
$Y= Y^+ + Y^-$ the separation into selfdual and antiselfdual
pieces. This is the projection to $(3,1)$, and $(1,3)$,
respectively.
 We claim that:
 \eqn\dimfouriden{ {
(\tanh Y)^+\over 1- \pf(\tanh Y)} = \half \tanh (2 Y^+) }

\noindent {\it Proof:} We first note that \dimfouriden\ is $SO(4)$
invariant. For $Y$ real antisymmetric there is always an SO(4)
rotation that skew diagonalizes it, so $Y = y_{12} T_{12} + y_{34}
T_{34} $, where $T_{ij} := e_{ij} - e_{ji} $, and $e_{ij}$ are
matrix units.  Then
\eqn\selfskew{ Y^+ = {1 \over 2} (y_{12} + y_{34}) (T_{12} +
T_{34}). }
In this skew diagonal form one easily checks \dimfouriden\ by
direct computation. Note that if $Y$ is skew diagonal then
\eqn\skewdiag{ \tanh Y = \tan( y_{12}) T_{12} +\tan (y_{34})
T_{34}. } Now we use the addition formula for tangents.
$\spadesuit$

Now, the solution to the equations
\eqn\pfrspin{ Y^{ij} \gamma_{ij} \eta =k \eta }
where $k$ is a constant and $\eta$ is of negative chirality, is
that $Y= Y^+ + Y^-$ where $Y^+$ is a constant in the stabilizer
subgroup $h=u(1)$.  (This constant can be  expressed in terms of
$k$) while the component $Y^-$ is an arbitrary function of
spacetime. Thus, it follows from \dimfouriden\ and $M=\e \, \tanh
(Y)$\foot{The factor of $\e$ arises because we have kept
$M_{\mu\nu}$ fixed while scaling the flat metric by $\e$. This
rescales $M_{\mu\nu}$ by $\e$. } that
 \eqn\deformpfr{
{\xi^{-1}(F+B)^+ \over 1 -\xi^{-2} \pf(F+B)} = {1\over 2} \tanh
(2Y ^+) = {\rm const.},} where we have introduced the parameter
\eqn\defofxi{ \xi = { \epsilon \over 2 \pi \alpha'}.}

The constant in \deformpfr\ is evaluated by going to infinity
(recall that we are now in a noncompact situation). As in \sw, we
will consider configurations in which $F \rightarrow 0$ at
infinity and $B$ is constant, corresponding to localized
instantons. This fixes the gauge freedom relating $F$ and $B$
completely, and we obtain:
\eqn\defeqfour{ {(F+B)^+ \over \pf(F+B)-\xi^2}={B^+ \over
\pf(B)-\xi^2}.} We can now compare this equation with what we
found in section 3. In order to do this, we have to specify the
appropriate complex structure. In the complex structure induced by
the reduced holonomy in \twofourag\ or \swfour, the $B$ field is a
$(1,1)$ form. Therefore, by choosing a complex structure for
\defeqfour\ in which $B$ is also of type $(1,1)$, we get again
\twofour.

\subsec{Deformation of the Hermitian Yang-Mills equations: $p=5$
and $p=7$}

We will now analyze the equation \aneq\ on $\IR^6$. This equation
says that the infinitesimal rotation by $Y$ preserves $\epsilon$
up to a constant. The covariantly constant spinor breaks the local
frame group  $SO(6) \rightarrow SU(3)$ at every point, and chooses
a complex structure. Relative to this complex structure we have
the deformed equations:
\eqn\twozero{ \eqalign{&Y^{2,0} = 0,\cr
 &J^{m \bar n} Y_{m\bar n} = k,\cr}  }
where $k$ is a constant.

Now we can analyze the meaning of the equations for $M/\epsilon =
\tanh Y$ along the lines of the previous subsection. $Y$ is now a
$ 6 \times 6$ antisymmetric matrix. Without loss of generality we
can take the local frame components of $J$ to be given by
$J=T_{12} + T_{34} + T_{56}$, so that $U(3)\subset SO(6)$ is
defined by $\{ A: A J = J A, A\in SO(6)\} $. Under this embedding
the antisymmetric rep ${\bf 15}$ of $SO(6)$ decomposes as:
\eqn\decomp{ {\bf 15} = {\bf 1} \oplus {\bf 3} \oplus {\bf 3^*}
\oplus {\bf 8} .} The matrix $Y$ decomposes then as: \eqn\ydec{
 Y = Y_1 + Y_3 + Y_{3^*} + Y_8.
} The equations \twozero\ say that $Y_3=Y_{3^*} = 0$ and $Y_1$ is
a constant times the identity, while $Y_8$ is an arbitrary
undetermined function.

The matrix $Z:=\tanh Y$ is also real antisymmetric and hence
decomposes as $Z =Z_1 + Z_3 + Z_{3^*} + Z_8$. The $Z_i$'s are
nonlinear functions of the $Y_i$'s. However, if $Y_3=0$ then
$Z_3=0$. This follows because in the power series representation
of $Z$, if $Y$ has zero triality then $Z$ has zero triality. The
conclusion of this analysis is that $Y^{2,0} =0 $ implies $M^{2,0}
=0$. Thus, ``most'' of the 6d Hermitian Yang Mills equations are
undeformed.

Moreover, if $Y_3=Y_{3^*} = 0$ then we have the identity:
\eqn\sixdeeiden{ { (\tanh Y)_1 - {1 \over 3} \pf(\tanh Y) J \over
1- \half \pf(\tanh Y){\rm  Tr} (J (\tanh Y)^{-1} ) } = {1 \over 3}
\tanh (3 Y_1). }
The left hand side defines a nonlinear function of $M/\epsilon=
\tanh Y$. Evaluating the constant by going to infinity we get a
deformation of the 6d Hermitian Yang-Mills equation, which can be
written as follows: \eqn\defsix{ {\xi^2(\CF \wedge J^2)/2!-\CF^3/3!
 \over \xi^2 J^3/3! - (\CF^2\wedge J)/2! } =
{\xi^2(B \wedge J^2)/2!-B^3/3!
 \over \xi^2 J^3/3! - (B^2\wedge J)/2!.}}
In this equation, $\CF=F+B$, we have assumed that $B$ is of type
$(1,1)$, and we have used the fact that $\CF^{2,0}=0$. Notice that
all the terms in \defsix\ are proportional to the volume form,
$J^3/3!$, and the quotient of forms should be understood as a
quotient of the corresponding scalars. \defsix\ can can also be
written as \eqn\comp{ {\biggl[ {\rm e}^{J} \sin (\xi^{-1}\CF)
\biggr]^{\rm top} \over \biggl[ {\rm e}^{J} \cos
(\xi^{-1}\CF) \biggr]^{\rm top}}= {\biggl[ {\rm e}^{J} \sin
(\xi^{-1}B) \biggr]^{\rm top} \over \biggl[ {\rm e}^{J} \cos
(\xi^{-1}B) \biggr]^{\rm top}},} where the superscript $\rm top$
means that we take the top form in the expansion, which in complex
dimension $n$ has degree $2n$.  Finally, notice that \defsix\
agrees with \sixsix\ (if, as we explained in the four-dimensional
case, $B$ is of type $(1,1)$). Also notice that the above deformed
equation solves \sixeight\ provided that the embedding is
holomorphic: \sixeight (c) holds if $M^{2,0}=0$, and, since
$f^*(\Omega)=0$, \sixeight (a) reduces to \defsix\ (for the value
of $\theta$ fixed by the behavior at infinity).

Equation \twozero\ straightforwardly generalizes to the next case
$p=7$ and further analysis gives as before $M^{2,0} =0$ (by virtue
of the decomposition under the embedding of $U(4)$ into $SO(8)$: $
{\bf 28} = {\bf 1} \oplus {\bf 6} \oplus {\bf 6^*} \oplus {\bf 15}
$ and four-ality). Moreover, we get again \comp, where the top
form has degree $8$. Explicitly, we find: \eqn\eight{ {\xi^3(\CF
\wedge J^3)/3!-\xi(\CF^3 \wedge J)/3!
 \over \xi^4 J^4/4! - \xi^2(\CF^2\wedge J^2)/(2!)^2 +\CF^4/4! } =
{\xi^3(B \wedge J^3)/3!-\xi(B \wedge J)/3!
 \over \xi^4 J^4/4! - \xi^2(B\wedge J^2)/(2!)^2 +B^4/4! }
.} In a manifold of $SU(4)$ holonomy, the equation for the
deformed instanton are also $M^{2,0}=0$ and \eight, with the only
difference that the constant in the right hand side is evaluated
in terms of the topological data, as in \keqs.

\subsec{$G_2$ and ${\rm Spin}(7)$}

As a further application let us consider the possible deformation
of the $G_2$-instanton  equations. Here $Y$ is in the ${\bf 21}$
of ${\rm Spin}(7)$ and a nonzero constant spinor $\chi$ determines
an embedding of $G_2$ into ${\rm Spin}(7)$. For example, from
$\chi$ we get $\gamma_{ijk} \chi = i \Phi_{ijk}\chi$ as in \ident,
and the $G_2$ subgroup is the subgroup of ${\rm Spin}(7)$
preserving $\Phi_{ijk}$. Now $Y= Y_7 + Y_{14}$. The equations
state that $Y_7=0$. This is equivalent to $(\tanh Y)_7 = 0 $
because tensor products of the ${\bf 14}$ never produce the ${\bf
7}$. (One way to see this is to note that weights in the tensor
products of the ${\bf 14}$ are always integer sums of root vectors
of $G_2$, so one can never produce the weights of the ${\bf 7}$
this way.)

In exactly the same way we find that the ${\rm Spin(7)}$ equations
are undeformed. That is, the component of $M$ in the ${\bf 7}$ of
${\rm Spin(7)}$ is zero.

\newsec{Relations to noncommutative instanton equations}

In this section, we will analyze the deformed instanton equations
of the previous section, focusing on their limiting behavior. In
this way we will recover and generalize some of the results in
\sw.

In the deformed equations of the previous section there are two
parameters, $\alpha'$ and $\epsilon$, which are combined into the
parameter $\xi$ defined in \defofxi.  We now consider several
different limiting behaviors. One important limit is to take
$\alpha' \rightarrow 0$ while keeping $\epsilon$ fixed (and
therefore $\xi \rightarrow \infty$). This is the usual zero slope
limit. Another interesting limit is discussed in \sw\cornalba.  In
\sw, Seiberg and Witten have shown that the relation of string
theory to noncommutative geometry appears in the double scaling
limit $\epsilon \rightarrow 0$, $\alpha' \sim \epsilon^{1/2}
\rightarrow 0$. In this limit (that will be called the
Seiberg-Witten limit), $\xi \sim \epsilon^{1/2} \rightarrow 0$.
Finally, we can consider the limit in which the $B$ field is set
to zero. We will now analyze these limits in the different
situations that we have considered.

\subsec{Instanton equation in four dimensions}

We will first analyze the deformed instanton equation \defeqfour.
In the zero slope limit, $\xi \rightarrow \infty$, and we get the
usual instanton equation $F^+=0$. For $B=0$, the equation reduces
again to the instanton equation, even for finite $\xi$. This is in
agreement with the observations in \sw, section 2.3. In order to
make contact with the deformed instanton equations that correspond
to noncommutative instantons, we have to take the Seiberg-Witten
limit $\xi \rightarrow 0$. We find, \eqn\sewq{ {(F+B)^+ \over
\pf(F+B)}={B^+ \over \pf(B)},} which is precisely the equation
(4.45) of \sw. In fact, our equation \defeqfour\ can be regarded
as a one-parameter deformation of the Seiberg-Witten equation,
where the deformation parameter is $\xi$. It is also illuminating
to write
\defeqfour\ in the open string frame, following the discussion in \sw. The
open string metric in the zero-slope limit is given by:
\eqn\open{ G_{ij} = -{(2 \pi \alpha')^2 \over \epsilon}
(B^2)_{ij},} which can be obtained from the vierbein \eqn\vierb{
E=-{\epsilon^{1/2} \over 2 \pi \alpha'} B^{-1}} as $G=(E
E^t)^{-1}$. In the open string frame, one has: \eqn\rels{
\eqalign{ F^+&={B^+ \over 4 {\rm Pf}(B)}
\epsilon^{ijkl}B_{ij}F_{kl} -({\rm Pf}(E))^{-1} E^t F_G^+ E,\cr
B^+&=({\rm Pf}(E))^{-1} E^t B_G^+ E,\cr} } where $F_G^+$, $B_G^+$
denote the self-dual projections in the open string metric \open.
Taking   into account that $\theta^{ij}= (B^{-1})^{ij}$, we find
the equation:
\eqn\defotwo{ F^+_G = { 1\over 4} { 1 \over 1-\xi^2 {\rm
Pf}(\theta)}(G\theta^+_G G) \bigl( {\widetilde F}^{ij}F_{ij} + 2
\xi^2{\rm Pf}(\theta){\widetilde F}^{ij} \theta^{-1}_{ij} \bigr),
}
where
\eqn\nottt{ \widetilde F^{ij}= {1 \over 2}{\epsilon^{ijkl} \over
{\sqrt {{\rm det} G}}}F_{kl}.}
\defotwo\ can be regarded as a two-parameter deformation of
the usual instanton equation, where the parameters are now
$\theta$ and $\xi$.

One can study spherically symmetric  solutions of \defotwo\ along
the lines of equations (4.56) to (4.62) of \sw. This exercise is
subject to the   criticism (discussed in \sw) that the instanton
equations are not valid near $r \rightarrow 0$ because the fields
are varying too rapidly there. One finds a one parameter family of
solutions interpolating between $h\sim 1/r^2 $ and $h\sim 1/r^4$.
It might be interesting to investigate solutions to the nonabelian
generalizations of these equations, because these might be
nonsingular.

 \subsec{Hermitian Yang-Mills equations}

We now study an analogous  deformation of the
  Hermitian Yang-Mills equations using the equations
   we have found above for   $p=5,7$.
First of all, in the zero slope limit $\xi \rightarrow \infty$ we
recover the ordinary Hermitian Yang-Mills equations, $F^{2,0}=0$,
$F\wedge J^{p-1 \over 2} =0$, as expected (we assume that $B$
is of type $(1,1)$). On the other hand, when $B=0$ and $\xi$ is
finite we do not recover these equations. Rather \eqn\bzero{
\biggl[ {\rm e}^{ J} \sin (\xi^{-1} F) \biggr]^{\rm top}=0,}
is still a deformation of Hermitian Yang-Mills. For example, for
$p=5$ we find \eqn\pfivebzero{ \xi^2{J^2\over 2!} \wedge F
-{1 \over 3!} F^3=0.} If the reasoning of \sw\ extends to this
case, then we should have a one-to-one correspondence between
solutions of \bzero\ and solutions of the usual Hermitian
Yang-Mills equations in six dimensions.

In the Seiberg-Witten limit $\xi\rightarrow 0$ we get the
equations:
\eqnn\swlim $$ \eqalignno{ &{\CF^{p-1 \over 2} \wedge J \over
\pf (\CF)} ={B^{p-1 \over 2} \wedge J \over \pf (B)},& \swlim
(a) \cr &\CF^{2,0}=0.& \swlim (b)\cr} $$
Using that $\phi^{p+1\over 2} = ({\rm Pf (\phi)})J^{p+1\over
2}$, $\phi^{p-1\over 2}\wedge J= {\rm Pf
(\phi)}\Tr(J\phi^{-1})$, where $\phi$ is a $(1,1)$ form, we
obtain:
\eqn\defffs{\Tr \Biggl[ J \biggl({1 \over F+B} - {1 \over
B}\biggr) \Biggr]=0.}
To compare with the non-commutative YM equations we should go to
the open string frame and recall that, ignoring the terms
involving derivatives in $F$, one has \sw:
\eqn\fhat{{\hat F}= {1 \over 1+F\theta}F = - B\biggl({1 \over B+F}
- {1 \over B}\biggr)B,}
where $\hat F$ is the field strength in the non-commutative
geometry defined in \sw. Using the vierbein \vierb, we see that
the equation
\defffs\ is equivalent to
\eqn\opend{\Tr (J_G {\hat F}) =0.}
Therefore, recalling that $B$ is of type $(1,1)$, we arrive at the
non-commutative Hermitian Yang-Mills equations:
\eqn\ncinst{\eqalign{{\hat F}^{2,0} &= 0, \cr {\hat F} \wedge
J ^{p-1\over 2} &=0.}}

It should be stressed that the formulae mapping to noncommutative
Yang-Mills theory used above apply to constant fieldstrengths $F$
of rank one, and moreover to  backgrounds with constant $B$.
Nevertheless, our BPS conditions apply to nonconstant $B$ and $F$,
and admit natural nonabelian generalizations, so it would be nice
to establish the equivalence between \swlim{a,b} and \ncinst\ in
the more general setting.  In particular, it would be interesting
to see if there is still a map to the noncommutative geometry
defined using the $*$ product of \kontsevich.

\newsec{Relations to previously studied nonlinear deformations of the instanton
equations }

The nonlinear deformations of instanton equations we have found
from $\kappa$-symmetry and BPS conditions are closely related to
some nonlinear instanton equations which have been previously
studied.

First, the equation \comp\  has some similarities to the nonlinear
deformations of the Hermitian Yang-Mills equations studied in
\leung. Leung introduced his equations to study the relation of
Gieseker and Mumford stability of holomorphic vector bundles.
Among Leung's results are some results that suggest that there
should be a 1-1 correspondence between the solutions of the
deformed and undeformed equations. This is certainly consistent
with the change of variables discussed by Seiberg and Witten.
Indeed, it suggests that their change of variables might be useful
in studying stability of holomorphic vector bundles.

Second, it is worth pointing out that  the equations \pfivebzero\
together with $F^{2,0}=0$ are just the equations of motion of the
``chiral cocycle theories'' studied by Losev et. al. in \clash.
These are theories of type $(1,1)$ connections on holomorphic
bundles governed by actions formed from Bott-Chern classes. The
chiral cocycle Lagrangians $\CL_n$ exist for complex manifolds
$X_n$ of any complex dimension $n$. They are constructed using
Bott-Chern forms and are functionals of a gauge field $A$
satisfying $F^{0,2}=0$. They have equation of motion:
\eqn\chircoc{ \delta \int_{X_n} \CL_n[g] = (\bar\p( g^{-1} \p
g))^n = 0 } where $A^{0,1} = - \bar \p g g^{-1} $ and $g\in {\rm
GL}(N,\IC)$. Therefore, using the Lagrangian \eqn\chicocii{
\sum_{k=0}^n a_k \int_{X_n}  J^{n-k} \CL_k[g] } for suitable
coefficients $a_k$ we can reproduce equations \comp\ above and
their limits \bzero\swlim. This connection is potentially useful
because, as discussed at length in \clash, the theories are
partially solvable using higher dimensional current algebra and
higher dimensional analogues of the ``$bc$-systems'' of 2D
conformal field theory. One wonders if the higher-dimensional
fermionization described in \clash\ could be useful in this regime
of string theory.

Moreover, the equations of \leung\ and of \clash\ both admit
natural {\it nonabelian} generalizations. The correct formulation
of a nonabelian Born-Infeld theory is a problem which has been
partially, but not fully solved \tseytlinrev. One also wonders if
the nonabelian chiral cocycle equations will be the equations for
BPS configurations of nonabelian Born-Infeld theories. If this is
the case then the connection could be very rich for mathematical
physics, providing natural nonlinear deformations of
Yang-Mills-Higgs systems.

\newsec{Kodaira-Spencer theory and the M5-brane}

The above analysis can also be applied to the $\kappa$-symmetries
of the $M2$ and $M5$ branes. In the case of the $M2$ brane, the
analysis has already been done in \bbs. Since  the only
worldvolume fields are scalars there is no nontrivial rotation of
the $\Gamma$ operator. This is consistent with the fact that we
found no interesting deformations for the case of $D1$ and $D2$
branes.

The situation for the M5 brane, on the other hand, is much
more nontrivial. Supersymmetric configurations on the M5
brane have been studied in \hlw\glw\gauntlett. Here we
focus on M5 instantons with none of the 5 normal bundle
scalars activated. Since we are working with
instantons we must decide on a formulation of the 5brane theory,
as well as a continuation of that theory to Euclidean space.
Since we are interested in on-shell configurations
we restrict attention to the purely on-shell and  covariant
formulation of \hsw\hswi. In this theory one uses a self-dual
3-form $h_{\mu\nu\rho}$ nonlinearly related to the field strength
of the 2-form potential, $H_{\mu\nu\rho}$. The latter
fieldstrength satisfies the Bianchi identity $dH \propto f^*(G_4)$
where $G_4$ is the $M$-theory 4-form fieldstrength.

In the formulation of \hswi\ one begins (in Minkowski space) with
a real self-dual 3-form $*h = h$. The nonlinear equation of motion
for $h$ is
\eqn\heom{ \eqalign{ \CM^{\mu\nu } \p_\mu h_{\nu\lambda\rho } & =
0 \cr \CM_\mu^{~~ \nu} & = \delta_\mu^{~~ \nu} - 2
h_{\mu\rho\lambda } h^{\nu \rho \lambda}. \cr} }
The $\Gamma$ operator defining $\kappa$-symmetry transformations
is   simply given by
$$\Gamma = \Gamma_{(0)}(1- {1 \over 2\cdot 3!} h_{\mu\nu\rho}
 \gamma^{\mu\nu\rho} ). $$
The crucial nonlinear relation of $h$ to the fieldstrength $H$ of
the 2-form potential is, according to \hswi, given by \eqn\nonlin{
\eqalign{ H_{\mu\nu\rho} & = (\CM^{-1})_{\mu}^{~~\lambda}
h_{\lambda \nu \rho} .\cr} }

%
%

While $H$ satisfies a simple Bianchi identity $dH=0$ (in a
background with $G_4=0$) the self-duality condition and the
$\Gamma$ operator are complicated nonlinear functions of $H$.
Indeed, we will regard the relation of $h$ to $H$ as quite
analogous to that between $Y$ and $M$, explored extensively in the
previous sections. In particular, we have learned from our
previous results that, while the equations for supersymmetric
brane configurations are complicated nonlinear equations on $M$,
they become much simpler in terms of $Y$. An analogous phenomenon
proves to be the case in the M5 theory.

Accordingly, let us examine the conditions on $h$ for a
supersymmetric M5 instanton.
 We will continue to Euclidean
 space by relaxing the reality condition on $h$ and taking
\eqn\hdul{*h = - i h. }
 On a K\"ahler manifold this implies that
$h$ is of the form
 $h= h^{3,0} + h^{2,1} + h^{1,2}$ where $h^{2,1}$ is
 in the image of $J\wedge$ and $h^{1,2}$ is in the
 kernel of $J\wedge$.
We take the equation of motion on a curved Euclidean manifold $X$
to be:
\eqn\heomeuc{ \CM^{\mu\nu } \nabla_\mu h_{\nu\lambda\rho } = 0. }
This implies $dH=0$ \hswi . Finally, we can continue $\Gamma$ to
Euclidean space by taking
 \eqn\eucgammaop{
\Gamma = \mp i \Gamma_{(0)}(1- {1 \over 2\cdot 3!} h_{\mu\nu\rho}
 \gamma^{\mu\nu\rho} ).
 }
The condition \hdul\ on $h$ guarantees that $\Gamma^2=1$.

We  now take $X$ to be a Calabi-Yau 3-fold and look for on-shell
field configurations $h$ such that there are covariantly constant
spinors with $\Gamma \eta = \eta$.
 Choosing the lower sign in \eucgammaop\
 we find no   condition on
 $h$. Choosing the upper sign we find the general solution
 \eqn\gsol{h = c \Omega + \chi^{1,2}}
 where $c$ is a constant, $\chi^{1,2}$ is of type $(1,2)$,
 and $J\wedge \chi^{1,2}= 0$, or
 equivalently $g^{m \bar n} \chi^{1,2}_{m \bar n \bar p } =0$.
It is now useful to define the variable
\eqn\mdef{{\mu_m}^{\bar n} := \half  \Omega_{mpq}\chi^{\bar n
pq}.}
 The condition $J\wedge
\chi^{1,2}= 0$ implies \eqn\mnf{\mu_{mn}=\mu_{nm}.}
Now we examine the implications of \heomeuc. This equation has
$(2,0)$, $(1,1)$, and $(0,2)$ components. We find that the $(2,0)$
component is identically satisfied thanks to $\nabla \Omega = 0 $.
The $(0,2)$ component becomes
\eqn\ohtwo{g^{m \bar n} \nabla_{\bar n} \chi_{m \bar p \bar q} - 4
\chi^n_{~~r \bar s} \chi^{m r \bar s} \nabla_n \chi_{m \bar p \bar
q} = 0  }
which is a deformation of the standard gauge fixing condition
$\p^\dagger \chi^{1,2} = 0$ of Kodaira-Spencer theory,  where
$\p^\dagger: \Omega^{1,2} \rightarrow \Omega^{0,2}$. Finally,
using \mnf\ repeatedly the $(1,1)$ component of \heomeuc\ becomes
{\it precisely} the Kodaira-Spencer equation
\eqn\rgp{\partial_{[m} {\mu_{n]}}^{\bar p}-8 c {\mu_{[m}}^{\bar
q}\partial_{\bar q} \mu_{n]}^{~~ \bar p}=0,}
  for a finite
deformation $\mu $ of the complex structure on $X$, as long as
$c\not=0$.

The problem we face at this point is that our three equations for
$\mu$ (or $\chi$), \mnf\ohtwo\ and \rgp, are potentially
overdetermined, hence it is not clear that they have solutions. We
conjecture that solutions in fact do exist, and that on a
Calabi-Yau manifold they are in one to one correspondence with the
solutions to the standard Kodaira-Spencer equations. We will now
give some partial evidence for this.

The Kodaira-Spencer equation has been explicitly solved on a
Calabi-Yau manifold by Tian and Todorov in \tian\todorov. The
first step in doing this is to set up a perturbative procedure to
solve the equation. We start from the ansatz: \eqn\pert{ \mu =
\sum_{n=1}^{\infty} \epsilon^n \mu^{(n)}, } where $\epsilon$ is a
formal parameter. We will denote by $': \Omega^{(p,0)} (\wedge^q
{\overline T_X}) \rightarrow \Omega^{(p, 3-q)}$ the contraction
with ${\overline \Omega}$, so that $\chi =\mu'$. Making a
convenient choice of $c$, the Kodaira-Spencer
equation at $n^{th}$
order is given by:
\eqn\kspert{
\partial \mu^{(n)} + {1\over 2} \sum_{i=1}^n [\mu^{(i)},\mu^{(n-i)}]=0.}
The resulting equations can be recursively solved in the gauge
$\p^{\dagger}\chi=0$. We will denote a solution in this gauge by
$\chi_{TT}$. At first order, one finds that $\chi_{TT}^{(1)}$ is
harmonic. At second order, and using that $[A,B]'={\overline \partial}(A\wedge
B)'$, the solution is given by: $$
\chi_{TT}^{(2)}=-\p^{\dagger}{1\over 2\Delta_{\partial}}{\overline
\partial} (\mu_{TT}^{(1)} \wedge
\mu_{TT}^{(1)})'. $$ One can in fact find an explicit solution
$\chi_{TT}$, constructed in a recursive way, which satisfies the
gauge condition $\partial^{\dagger}\chi_{TT}=0$ and also
${\overline
\partial}\chi_{TT}=0$. This solution is
given, at $n$th order, by \tian\todorov: \eqn\general{
\chi_{TT}^{(n)}=-\p^{\dagger}{1\over
2\Delta_{\partial}}\sum_{i=1}^n {\overline \partial} (\mu_{TT}^{(i)}
\wedge \mu_{TT}^{(n-i)})',} and $\chi^{(1)}_{TT}$ is any harmonic
$(1,2)$ form in the Calabi-Yau. A remarkable fact is that this
solution satisfies automatically the extra equation \mnf, or
equivalently, $J\wedge \chi_{TT}=0$. This can be proved inductively
as follows: Consider the $(2,3)$-form $J\wedge \chi_{TT}^{(1)}$.
Since $\chi_{TT}^{(1)}$ is harmonic, using the Hodge identity
$[J\wedge,
\p^{\dagger}]=i{\overline \p}$ we can easily prove that this
$(2,3)$-form is also harmonic. But $h^{2,3}=0$ on a Calabi-Yau, so
$J\wedge \chi_{TT}^{(1)}=0$. This proves \mnf\ at first order.
Let's now assume that $J\wedge \chi_{TT}^{(i)}=0$ for $i=1, \dots,
n-1$. Then \eqn\next{ J\wedge \p^{\dagger}{1\over
2\Delta_{\partial}} {\overline \partial} (\mu_{TT}^{(i)} \wedge
\mu_{TT}^{(n-i)})'=\p^{\dagger} {1\over 2\Delta_{\partial}}
{\overline \partial} (J\wedge (\mu_{TT}^{(i)} \wedge
\mu_{TT}^{(n-i)})'),} and $J\wedge (\mu_{TT}^{(i)} \wedge
\mu_{TT}^{(n-i)})'$ is easily seen to be zero after using the
definition of $'$ and the induction hypothesis.

Our equations involve the deformed gauge condition \ohtwo\  rather
than the one used in the proof of the  Tian-Todorov theorem. Since
we are just changing the gauge, we can try to find a solution to
our equations of the form $\chi^{(n)} = \chi_{TT}^{(n)} +\delta
\chi^{(n)}$, where $\chi_{TT}^{(n)}$ is the explicit solution
\general, and in such a way that \mnf\ is still true. The first
step in doing this is to rewrite \ohtwo\ as follows. We introduce
the determinant of $\mu$, that we will denote by $\det \mu :=
\det_{i, \bar j}  \mu_i^{~~ \bar j}$. If we choose $\vert\vert
\Omega\vert\vert^2=1$, we find that \ohtwo\ can be written as
\eqn\stepfour{ \nabla^m \mu_{mk} = - {1 \over  8c} \biggl(\nabla_k
- 8 c \mu_{km} \nabla^m\biggr) \log[1+ 64 c  (\det \mu) ]. }
We can also write \stepfour\ as \eqn\stepfive{ (\partial^{\dagger}
\chi)_{\bar p \bar q} = ({\overline
\partial}^{\dagger} (f{\overline \Omega}))_{\bar p \bar q} -8c
\chi_{m \bar p \bar q} \nabla^m f,} where $f=- {1 \over  8c}
\log[1+ 64 c (\det \mu)]$. At first and second order, the solution
to our equations is just given by $\chi_{TT}^{(n)}$, $n=1,2$,
since the deformation of the gauge fixing condition is cubic in
$\chi$. At third order, the gauge fixing condition becomes:
\eqn\gaugethree{ (\partial^{\dagger} \chi^{(3)})_{\bar p \bar q}
=(\partial^{\dagger}\delta \chi^{(3)})_{\bar p \bar q}
=({\overline \partial}^{\dagger} (f^{(3)}{\overline
\Omega}))_{\bar p \bar q},} where $f^{(3)}$ is the third order
term in $\mu$ (and in fact involves only $\mu^{(1)}$). A change of
gauge at third order simply means that $\delta \chi^{(3)}=\partial
\nu$, where $\nu$ is a $(0,2)$ form that satisfies: \eqn\nucond{
\Delta_{\partial} \nu = {\overline \partial}^{\dagger}
(f^{(3)}{\overline \Omega}).} We can then write \eqn\del{ \delta
\chi^{(3)}=\p {1\over \Delta_{\p}} {\overline
\partial}^{\dagger}(f^{(3)} {\overline \Omega}),}
and it is easy to check (using again the Hodge identities) that
$J\wedge \delta \chi^{(3)}=0$. Therefore, the perturbation of the
solution \general\ induced by the deformation of the gauge
condition preserves \mnf\ at third order. Unfortunately, the
procedure becomes cumbersome for higher orders and we have not
been able to check it for the next terms in the perturbative
series. We conjecture, however, that the equations with the new
gauge fixing condition can be solved in the way that we have
sketched. (In arranging a full proof it might help to notice that
the right hand side of \stepfour\ involves the deformed
holomorphic derivative $\nabla_k - 8 c \mu_{km} \nabla^m$.)  In
particular, we conjecture that the solutions to our equations are
in one to one correspondence with the solutions to the
Kodaira-Spencer equation with the usual gauge fixing, and in such
a way that \mnf\ is satisfied.

Our result is relevant to the problem of computing nonperturbative
corrections for $M$-theory on a Calabi-Yau. These will involve a
weighted sum over all configurations of wrapped fivebranes with
supersymmetric $H$ fields turned on. The preceding relates such
$H$ fields to points in the moduli space of complex structures on
the Calabi-Yau. $H$ is subject to a quantization condition which
restricts the sum to rational points reminiscent of those arising
from the attractor equations \refs{\fks,\as}.

Our  result also establishes a very direct relation between the
M5-brane and Kodaira-Spencer theory. Connections between the M5
theory and Kodaira-Spencer theory have been discussed before. In
particular, in \wittenbcov\ Witten related the quantization of the
phase space $H^3(X)$ to Kodaira-Spencer theory and the holomorphic
anomaly equation \bcov. (See \todorov\ for some recent progress.)
In \wittenfiveact\ Witten then connected the quantization of
$H^3(X)$ to the M5 theory. Closely related connections have been
explored in \dvvunpub. Nevertheless, we believe the above
connection is new. We hope it leads to further progress in
demystifying the $M$-theory fivebrane.

\bigskip
\centerline{\bf Acknowledgments }\nobreak
\bigskip

We would like to thank D. Gross, N. Seiberg, G. Tian and S.-T. Yau for
useful discussions on noncommutative geometry and deformations of
instantons.
 GM and AS  would like to acknowledge the hospitality of
the Aspen Center for Physics. MM would like to thank L.
\'Alvarez-C\'onsul and O. Garcia-Prada for pointing out reference
\leung. GM would like to thank R. Dijkgraaf, A. Gerasimov, A.
Losev and S. Shatashvili for discussions (c. 1997) of relations
between fivebranes and Kodaira-Spencer theory. This work was
supported by DOE grants DE-FG02-92ER40704 and DE-FG02-91ER40654.

\listrefs
\end